\documentclass[twocolumn,eqsecnum,showpacs,showkeys,amsmath,amssymb,nofootinbib,superscriptaddress,floatfix]{revtex4}

\usepackage{graphicx}
\usepackage{subfigure}
\usepackage{bm}

\makeatletter
\newcommand\erfc{\mathop{\operator@font erfc}\nolimits}
\def\slashchar#1{\setbox0=\hbox{$#1$}
   \dimen0=\wd0 \setbox1=\hbox{/} \dimen1=\wd1
   \ifdim\dimen0>\dimen1 \rlap{\hbox to \dimen0{\hfil/\hfil}} #1
   \else  \rlap{\hbox to \dimen1{\hfil$#1$\hfil}} / \fi}

\makeatother

\begin{document}
 
\title{Generalized parton distributions of the pion \\ in chiral quark
models and their QCD evolution\footnote{Supported by Polish Ministry of Science and Higher
Education, grants 2P03B~02828 and N202~034~32/0918, Spanish DGI and FEDER
funds with grant FIS2005-00810, Junta de Andaluc{\'\i}a grant
FQM225-05, and EU Integrated Infrastructure Initiative Hadron Physics
Project contract RII3-CT-2004-506078.
}}

\author{Wojciech Broniowski} \email{Wojciech.Broniowski@ifj.edu.pl}
\affiliation{The H. Niewodnicza\'nski Institute of Nuclear Physics,
PL-31342 Krak\'ow, Poland} \affiliation{Institute of Physics,
\'Swi\c{e}tokrzyska Academy, ul.~\'Swi\c{e}tokrzyska 15,
PL-25406~Kielce, Poland} \author{Enrique Ruiz Arriola}
\email{earriola@ugr.es} \affiliation{Departamento de F\'{\i}sica
At\'omica, Molecular y Nuclear, Universidad de Granada, E-18071
Granada, Spain} \author{Krzysztof Golec-Biernat}
\email{golec@ifj.edu.pl}  \affiliation{The
H. Niewodnicza\'nski Institute of Nuclear Physics, PL-31342 Krak\'ow,
Poland}\affiliation{Institute of Physics, Rzesz\`ow
University, PL-35959 Rzesz\'ow, Poland}

\date{6 December 2007}

\begin{abstract}
We evaluate Generalized Parton Distributions of the pion in 
two chiral quark models: the Spectral Quark
Model and the Nambu--Jona-Lasinio model with a Pauli-Villars
regularization. We proceed by the evaluation of double distributions
through the use of a manifestly covariant calculation based on the
$\alpha$ representation of propagators.  As a result polynomiality is
incorporated automatically and calculations become simple. In
addition, positivity and normalization constraints, sum rules and soft
pion theorems are fulfilled. We obtain explicit formulas, holding at
the low-energy quark-model scale. The expressions exhibit no 
factorization in the $t$-dependence. The QCD evolution of those parton
distributions is carried out to experimentally or lattice accessible
scales. We argue for the need of evolution by comparing the Parton Distribution Function 
and the Parton Distribution Amplitude of the pion to the available experimental and lattice data, 
and confirm that the quark-model scale is low, about 320~MeV.
\end{abstract}

\pacs{12.38.Lg, 11.30, 12.38.-t}

\keywords{generalized parton distributions, double distributions,
light-cone QCD, exclusive processes, chiral quark models}

\maketitle 

\section{Introduction}

Generalized Parton Distributions (GPD's) encode detailed dynamical
information on the internal structure of hadrons and have thus become
in recent times a major theoretical and experimental endeavor (for
extensive reviews see
e.g.~\cite{Ji:1998pc,Radyushkin:2000uy,Goeke:2001tz,
Diehl:2003ny,Ji:2004gf,Belitsky:2005qn,Feldmann:2007zz,Boffi:2007yc}
and references therein). They represent a natural interpolation
between form factors and quark distribution functions. Actually, while
form factors and distribution functions provide in a separate way the
spatial and momentum quark distributions in a hadron respectively,
GPD's provide a simultaneous phase-space description of the quark
hadron content as far as the position-momentum uncertainty relations
allow~\cite{Belitsky:2003nz,Ji:2004gf}. Experimentally, GPD's show up
in hard exclusive processes such as Deeply Virtual Compton Scattering
(DVCS) or hard electroproduction of mesons.  Factorization for hard
exclusive electroproduction of mesons in QCD was proved in
Ref.~\cite{Collins:1996fb}. Effects of the Regge exchanges to exclusive 
processes were investigated in Ref.~\cite{Szczepaniak:2007af}.

In the present paper we are interested in GPD's for the pion. Although
there are little chances of measuring them directly in experiment,
the pion GPD's are amenable to indirect experimental determination as
well as studies both on transverse~\cite{Burkardt:2001jg} as well as
Euclidean~\cite{Hagler:2007hu} lattices (for a combination of experimental and lattice-based
reconstruction in the proton case see, {\em e.g.},
Ref.~\cite{Ahmad:2007vw}). In addition, there are many
theoretical advantages for studying this quark-antiquark bound
state. In the first place, in the chiral limit where the current quark
masses vanish, the spontaneous breakdown of chiral symmetry of the QCD
vacuum generates pions as the zero modes. Their properties are
expected to be dominated by the broken chiral symmetry while
confinement effects are expected not to be crucial. Actually, finite
mass corrections to GPD's of the pion have been treated in the
standard~\cite{Diehl:2005rn} and partially-quenched~\cite{Chen:2006gg}
chiral perturbation theory, while the breakdown of the expansion
for small $x \sim m_\pi^2/(4 \pi f)^2 $ has been pointed out in
Ref.~\cite{Kivel:2007jj}.  Besides, compared to the nucleon there are
no spin complications for the pion case, and thus the study reduces to
two single scalar GPD function, one for each isospin combination.
Finally, the pion provides a useful framework to learn on the
interplay between the chiral symmetry and the light cone features,
since we are studying the behavior of the would-be Goldstone boson in
the infinite momentum frame.

Despite the intrinsic complexity of the GPD's, there is a number of
simple conditions which ultimately are consequences of the Poincar\'e
and electromagnetic gauge invariance and provide {\it a priori}
tests on the validity of theoretical calculations.  Proper support and
polynomiality restrictions on the GPD moments~\cite{Ji:1998pc} are
manifestations of the Lorentz invariance. We note that polynomiality is not 
satisfied in
light-front calculations. Double distributions do not suffer from this
problem~\cite{Radyushkin:1998es} (using the double
distributions is a way of {\it projecting} the Lorentz-violating term
onto the right space) although they require the so-called
$D$-terms~\cite{Polyakov:1999gs} to comply with the most general
polynomial allowed by the dimensional analysis.  Normalization
conditions and sum rules are a manifestation of the gauge invariance,
which at the quark level requires the correct implementation of the
electromagnetic Ward-Takahashi identities. The positivity
bound~\cite{Pire:1998nw,Pobylitsa:2001nt} underlines the Hilbert-space
quantum-mechanical probabilistic nature of pion light-cone wave
functions, and may impose relevant constraints on admissible
regularizations based mainly on subtractions of ultraviolet
divergences. Soft pion theorems based on PCAC relate GPD's to Parton
Distribution Amplitudes (PDA's)~\cite{Polyakov:1998ze}. On a
theoretical level, the amazing aspect of GPD's is that the constraints
that ought to be fulfilled {\it a priori} are so demanding and
intricate that it is extremely difficult to provide ans\"atze
fulfilling all of them simultaneously. This is why dynamical
calculations going beyond reasonable but admittedly {\it ad hoc}
parameterizations are particularly interesting and instructive. On the
other hand, dynamical models providing GPD's are also generating
mutually consistent parton distribution functions, parton distribution
amplitudes, and form factors.  Although this may appear a rather
trivial statement, it imposes demanding and tight constraints on
details of the calculation, and more specifically on the proper
handling of ultraviolet divergences based on the correct
implementation of electromagnetic and chiral Ward-Takahashi
identities. Even in the case of a simple hadron such as the pion in
the chiral limit, the above mentioned necessary conditions provide
powerful limitations and in some cases clash with well established
prejudices about the meaning and realization of a relativistic
bound-state wave functions in quantum field theory \cite{Davidson:1994uv,Weigel:1999pc,VanDyck:2007jt}.

In the present work we determine GPD's incorporating all the desirable
properties required by the symmetries in two chiral quark models, the
Nambu--Jona-Lasinio (NJL)~(for reviews see
e.g. \cite{Christov:1995vm,RuizArriola:2002wr} and references therein)
and the Spectral Quark model
(SQM)~\cite{RuizArriola:2001rr,RuizArriola:2003bs}, which essentially 
is a way of introducing regularization in such a way that the vector-meson 
dominance is built in. 
These field
theoretical models incorporate the spontaneous breaking of chiral
symmetry. Chiral quark models use the large-$N_c$ expansion at leading
order, where the observables are obtained with one-loop quark
diagrams. The present calculation extends previous calculations of
PDF's ~\cite{Davidson:1994uv,RuizArriola:2001rr,Davidson:2001cc} and
PDA's \cite{RuizArriola:2002bp, RuizArriola:2002wr}. Diagonal GPD in
impact parameter space in these models were considered
\cite{Broniowski:2003rp}. Our present GPD result reproduces
consistently all these particular cases. Recently, the Transition
Distribution Amplitude (TDA) \cite{Pire:2004ie,Pire:2005ax} has also been evaluated in
SQM~\cite{Broniowski:2007fs}.

There has been a number of calculations of GPD's and related
quantities of the pion within the framework incorporating chiral
symmetry. Early calculations of pion GPD's were done in an
instanton-inspired model characterized by a momentum dependent mass
function of a dipole form
\cite{Polyakov:1999gs,Anikin:2000th,Anikin:2000sb}, while PDF's 
were evaluated in the same model in \cite{Dorokhov:1998up,Dorokhov:2000gu}. The
crossing-related $2\pi$GPA was also evaluated in
Ref.~\cite{Polyakov:1998td} in the same instanton model disregarding
the momentum dependence of the quark mass, a valid assumption in the
limit of small instantons. In that limit, end point discontinuities
arise. A full consideration of poles in the complex plane in a
non-local version of the NJL model was described
in Ref.~\cite{Praszalowicz:2003pr}.  Generally, the nonlocality of the
quark mass function generated incorrect normalization, since as noted
later, PCAC should be properly incorporated~\cite{Bzdak:2003qe}, an
issue also emphasized more recently in Ref.~\cite{Noguera:2005cc}. A
rather interesting feature of Ref.~\cite{Bzdak:2003qe} is that
end-point discontinuities reappear after PCAC is incorporated,
even for momentum-dependent quark mass functions, against the widely
spread prejudice that they only arise for momentum-independent
masses. The OPE and duality aspects of GPD's have been discussed in
Ref.~\cite{Bakulev:2000eb}. Light-front calculations have been
undertaken in Ref.~\cite{Choi:2001fc,Choi:2002ic} for point couplings
with subsequent insertion of Gaussian wave functions, however the
approach violates polynomiality. Power-law wave functions and GPD's of
the pion proposed in Ref.~\cite{Mukherjee:2002gb} satisfy
polynomiality but violate positivity (see
\cite{Tiburzi:2002kr}). Studies paying particular attention to
polynomiality were first made by Tiburzi and Miller
\cite{Tiburzi:2002kr,Tiburzi:2002tq} who proceeded via double
distributions~\cite{Radyushkin:1998bz}. However, regularization in
these works was done by introducing momentum dependent form factors,
which makes them difficult to reconcile with the gauge invariance.
The model of Ref.~\cite{Bissey:2003yr} based on a pseudoscalar
pion-quark coupling does not incorporate chiral symmetry and does not
fulfill the momentum sum rule.  Noguera, Theu\ss{}l, and Vento carried
out a calculation in the NJL model based on the light-front
coordinates, where the fulfillment of polynomiality for non-vanishing
momentum transfer is not apparent analytically~\cite{Theussl:2002xp},
and in fact numerical integration was required to establish this
property. Our NJL results agree with that work, with the important
methodological difference that the double distributions, where
polynomiality is manifest, are used throughout. Moreover, our
regularization is somewhat different than in the model of
Ref.~\cite{Theussl:2002xp}, we also use the non-linear rather than
linear realization of the chiral field.

Despite the numerous model calculations of the GPD's, it remains to
date unclear what is their significance or impact on the
interpretation of actual experiments and/or lattice data. This is,
perhaps, why most calculations of the GPD's based on dynamical quark
models and going beyond just phenomenological parameterizations do not
address this issue. However, while experimental or lattice results
generate scale-dependent GPD's, embodying the well established
logarithmic scaling violations in QCD, it is notorious that models
generally produce scale independent functions. Thus, quark models
represent those distributions at a {\it given} low energy scale. It is
noteworthy that scaling violations can only be computed in the twist
expansion order by order. For instance, for the structure functions
$F(x,Q)$ with the Bjorken $x$ and momentum $Q$ one has
for the quark model
\begin{eqnarray}
F(x,Q) = F_0 (x) + \frac{F_2 (x)}{Q^2} + \dots,
\end{eqnarray}
while for QCD
\begin{eqnarray}
F(x,Q) = F_0 (x,\alpha (Q^2)) + \frac{F_2 (x,\alpha (Q^2))}{Q^2} + \dots,
\end{eqnarray}
where $F_n (x,\alpha(Q^2))$ are low energy matrix elements with computable
anomalous dimensions and depending logarithmically on the scale through 
the running coupling constant 
\begin{eqnarray}
&&\alpha(Q^2)=\frac{4\pi}{\beta_0 \log(Q^2/\Lambda^2_{QCD})} \\
&&\beta_0=\frac{11}{3}N_c-\frac{2}{3}N_f. \label{gambe}
\end{eqnarray}
In this work we take 
\begin{eqnarray}
\Lambda_{\rm QCD} = 226~{\rm MeV} \label{LambdaQCD}
\end{eqnarray}
and $N_c=N_f=3$.  The
matching conditions are taking at a given scale $Q_0$ order by order
in the twist expansion
\begin{eqnarray}
F_n (x) |_{\rm Model} = F_n ( x, \alpha (Q_0^2 ) )|_{\rm QCD} .\label{match}
\end{eqnarray}
A quite different issue is the operational definition of the low-energy 
reference scale $Q_0$. Here we will use along the lines of
previous
works~\cite{Davidson:1994uv,Davidson:2001cc,RuizArriola:2002bp,
RuizArriola:2002wr} the momentum fraction carried by the valence
quarks.  It turns out that in order to describe the available pion
phenomenology the initial scale $Q_0$ from the quark model must be
very low, around $320~{\rm MeV}$. At such low scale the perturbative
expansion parameter in the evolution equations is large,
${\alpha(Q^2_0)}/{2\pi}=0.34$, which makes the evolution very fast for
scales close to the initial value $Q_0$.

None of the previous chiral-quark-model studies of the 
genuine GPD's (off-forward non-diagonal) carried
out the QCD evolution, starting from the initial condition provided by
the models. The evolution is a major element of this work. As already
mentioned, it is also a crucial element if one wishes to compare the
model prediction to the data from experiments or lattice
simulations. At the moment these data are available only for the
forward diagonal parton distribution function of the pion (PDF), or the PDA.

We have taken an effort to separate formal aspects of the calculation
from the model-dependent technicalities of the regularization. That
way we simply achieve the desired features on general grounds, such as
the sum rules or the polynomiality conditions~\cite{Ji:1998pc}. We stress this
is achieved without a factorized form in the $t$ variable. In one of
the considered models (the Spectral Quark Model) the final results for
the GPD's can be written in terms of rather simple but non-trivial analytic formulas,
which allows for more insight into their properties.  We also show that
our GPD's satisfy the positivity bounds. In essence, all the known
consistency conditions and constraints are indeed satisfied in our calculation.

The outline of the paper is as follows: In Sect.~\ref{sec:def} we list
the definitions and properties of the GPD's. We introduce both the
asymmetric and symmetric kinematics, as well as define the isospin
projections. We give the generic quark-model expressions in terms of the
two- and three-point functions. Sections~\ref{sec:sqm} and
\ref{sec:njl} contain the results of the Spectral Quark Model and the
NJL model, respectively. We discuss the general need for the QCD evolution
of chiral quark models in Section \ref{sec:qcd-qm}, where we define
the matching condition in the light of phenomenological analyses, as well as
Euclidean and transverse lattice calculations.  In particular, we show
the evolved forward diagonal PDF of the pion and confront it with the
results of the E615 experiment at
Fermilab~\cite{Conway:1989fs}. This agreement is quite remarkable, but
sets the quark model momentum scale to very low values, $Q_0 \simeq
320$~MeV. Likewise we also discuss the evolved PDA as compared to the
E791 measurement~\cite{Aitala:2000hb} of the pion light-cone wave
function and to the lattice data. 
We show how the QCD evolution leads to vanishing of the PDF at $x=1$ 
and of PDA at $x=0,1$, which is the desired end-point behavior.
The LO QCD evolution of our 
genuine GPD's, based on the standard ERBL-DGLAP
equations, is carried out in Sect.~\ref{sec:evol}, where the obtained
quark-model initial conditions are evolved to higher momentum scales.
Finally, in Section \ref{concl} we draw our main conclusions.
The Appendices contain the technique of analyzing the GPD's through the
use of the $\alpha$ representation for the propagators. This method, first 
introduced in the context of structure functions in Refs.~\cite{Dorokhov:1998up,Dorokhov:2000gu},
allows for a manifestly covariant calculation and leads to simple
formal expressions for the basic two- and three-point functions
emerging in the analysis. The appearance of $D$-terms is manifest and natural in
this treatment, based solely on the Feynman diagrams in a conventional
way. We also list explicitly the basic two- and three-point functions of the two
considered model. We proceed via the double distributions, which leads
to a simple proof of polynomiality~\cite{Radyushkin:2000uy}.

\section{Definitions and quark-model expressions \label{sec:def}}

\subsection{Formalism for pion GPD}
\label{sec:formalism}

The kinematics of the process and the assignment of momenta (in the
asymmetric way) is displayed in Figs.~\ref{fig:diag} and
\ref{fig:diag2}.  For the pions on the mass shell we have, adopting
the standard notation
\begin{eqnarray}
&& p^2=m_\pi^2, \;\; q^2=-2p\cdot q=t,\nonumber \\ && n^2=0, \;\; p
\cdot n=1, \;\; q \cdot n=-\zeta. \label{kin}
\end{eqnarray}
Note the sign convention for $t$, positive in the physical region.

The leading-twist off-forward ($t \neq 0$) non-diagonal ($\zeta \neq
0$) generalized parton distribution (GPD) of the pion is defined as
\begin{eqnarray}
&& {\cal H}^{ab}(x,\zeta,t) = \int \frac{dz^-}{4\pi} e^{i x p^+ z^-}
\times \label{defGPD} \\ && \;\;\; \left . \langle \pi^b (p+q) | \bar
\psi (0) \gamma \cdot n \,T\, \psi (z) | \pi^a (p) \rangle \right
|_{z^+=0,z^\perp=0}, \nonumber
\end{eqnarray} 
where $0 \le \zeta$ and the $x$ variable, $-1+\zeta \le x \le 1$, is
defined in the {\em asymmetric} notation ({\em cf.}
Fig.~\ref{fig:diag}), $a$ and $b$ are isospin indices for the pion,
$T$ is the isospin matrix equal $1$ for the isoscalar and $\tau_3$ for
the isovector cases, finally $\psi$ is the quark field and $z$ is the light-cone
coordinate.  Explicitly,
the two isospin projections are equal to
\begin{eqnarray}
&& \delta_{ab}\,{\cal H}^{I=0}(x,\zeta,t) = \int \frac{dz^-}{4\pi}
e^{i x p^+ z^-} \times \\ && \;\;\; \left . \langle \pi^b (p+q) | \bar
\psi (0) \gamma \cdot n \psi (z) | \pi^a (p) \rangle \right
|_{z^+=0,z^\perp=0}, \nonumber \\ && i \epsilon_{3ab}\,{\cal
H}^{I=1}(x,\zeta,t) = \int \frac{dz^-}{4\pi} e^{i x p^+ z^-} \times
\label{defGPD01} \\ && \;\;\; \left . \langle \pi^b (p+q) | \bar \psi
(0) \gamma \cdot n \psi (z) \, \tau_3 | \pi^a (p) \rangle \right
|_{z^+=0,z^\perp=0}. \nonumber
\end{eqnarray} 
One can form the combinations termed the quark and antiquark GPD's of the pion, 
\begin{eqnarray}
\!\!\!\!{\cal H}_q(x,\zeta,t)&=&\frac{1}{2}\left ( {\cal
H}_{I=0}(x,\zeta,t)+{\cal H}_{I=1}(x,\zeta,t)\right ), \nonumber \\
\!\!\!\!{\cal H}_{\bar q}(x,\zeta,t)&=&\frac{1}{2}\left ( {\cal
H}_{I=0}(x,\zeta,t)-{\cal H}_{I=1}(x,\zeta,t)\right ). \label{defH}
\end{eqnarray}
From the general formulation it follows that ${\cal H}_q(x,\zeta,t)$
has the support $x \in [0,1]$, whereas ${\cal H}_{\bar q}(x,\zeta,t)$
the support $x \in [-1+\zeta,\zeta]$.  The range $x \in [0,\zeta]$ is
called the ERBL region, while $x \in [-1+\zeta,0]$ and $x \in
[\zeta,1]$ are the DGLAP regions, where the nomenclature refers to the
QCD evolution, see Sect.~\ref{sec:evol}.

In the {\em symmetric} notation, somewhat more convenient in certain
applications\footnote{In this paper we switch back and forth between
the two conventions, since explicit expressions are shorter in the
asymmetric notation, while some formal features are simpler to state
in the symmetric notation.}, one introduces
\begin{eqnarray}
\xi= \frac{\zeta}{2 - \zeta}, \;\;\;\; X = \frac{x - \zeta/2}{1 -
\zeta/2}, \label{xiX}
\end{eqnarray}
where $0 \le \xi \le 1$ and $-1 \le X \le 1$. Then 
\begin{eqnarray}
H^{I=0,1}(X,\xi,t)={\cal H}^{I=0,1}\left ( \frac{\xi + X}{\xi + 1},
\frac{2 \xi}{\xi + 1},t \right ).
\end{eqnarray}
with the symmetry properties about the $X=0$ point,
\begin{eqnarray}
H^{I=0}(X,\xi,t)&=&-H^{I=0}(-X,\xi,t), \nonumber \\
H^{I=1}(X,\xi,t)&=&H^{I=1}(-X,\xi,t). \label{eq:sym}
\end{eqnarray} 
The following sum rules hold:
\begin{eqnarray}
&&\int_{-1}^1 \!\!\!\!\! dX\, {H}^{I=1}(X,\xi,t) = 2 F_V(t), \label{norm} \\
&&\int_{-1}^1 \!\!\!\!\! dX\,X \, {H}^{I=0}(X,\xi,t) = \theta_2(t)-\xi^2 \theta_1(t), \label{norm2} 
\end{eqnarray}
where $F_V(t)$ is the electromagnetic form factor, while $\theta_1(t)$
and $\theta_2(t)$ are the gravitational form factors of the pion (see
Appendix~\ref{app:grav}) which satisfy the low energy theorem
$\theta_1(0) =\theta_2(0) $ in the chiral
limit~\cite{Donoghue:1991qv}. Sum rule (\ref{norm}) expresses the
electric charge conservation, while (\ref{norm2}) is responsible for
the momentum sum rule in deep inelastic scattering. Finally, for $X
\ge 0$
\begin{eqnarray}
{\cal H}^{I=0,1}(X,0,0) = q(X), \nonumber
\label{eq:q} 
\end{eqnarray} 
relating the distributions to the the pion's forward diagonal parton
distribution function (PDF), $q(X)$.

The {\em polynomiality} conditions~\cite{Ji:1998pc,Radyushkin:2000uy}
state that
\begin{eqnarray}
\int_{-1}^1 \!\!\!\!\! dX\,X^{2j} \, {H}^{I=1}(X,\xi,t) = \sum_{i=0}^j A^{(j)}_i(t) \xi^{2i}, \nonumber \\
\int_{-1}^1 \!\!\!\!\! dX\,X^{2j+1} \, {H}^{I=0}(X,\xi,t) = \sum_{i=0}^{j+1} B^{(j)}_i(t) \xi^{2i}, \label{poly}
\end{eqnarray}
where $A^{(j)}_i(t)$ and $B^{(j)}_i(t)$ are the coefficient functions
(form factors) depending on $j$ and $i$.  The polynomiality conditions
follow from basic field-theoretic assumptions such as the Lorentz
invariance, time reversal, and hermiticity, hence are automatically
satisfied in approaches that obey these requirements. Conditions
(\ref{poly}) supply important tests of consistency.  In our approach
the polynomiality will be demonstrated straightforwardly in an
analytic way through the use of double distributions, see
Appendix~\ref{sec:alpha}.

Another constraint for the GPD's, the {\em positivity
bound}~\cite{Pobylitsa:2001nt}, is derived with the help of the
Schwartz inequality and the momentum representation of the pion
light-cone wave functions. In the simplest form the constraint states
that (for $t \le 0$)
\begin{eqnarray}
|H_q(X,\xi,t)| \le \sqrt{q(x_{\rm in}) q(x_{\rm out})}, \;\;\;\;\; \xi
 \le X \le 1. \label{positiv}
\end{eqnarray}  
where $x_{\rm in}=(x+\xi)/(1+\xi)$, $x_{\rm
out}=(x-\xi)/(1-\xi)$.

The off-forward ($\bf{\Delta_ \perp} \neq 0$) diagonal ($\xi=0$)
GPD of the pion (we take $\pi^+$) can be written as 
\begin{eqnarray}
H(x,\xi = 0, - {\bf{\Delta}}_ \perp^2 ) = \int d^2 {\bf b} \, e^{i
{\bf{\Delta}}_\perp \cdot {\bf{b}} } \, q(x,{\bf b}).
\end{eqnarray} 
We use here
\begin{eqnarray}
&& q(x,{\bf b}) = \int
\frac{dz^-}{4\pi} e^{i x p^+ z^- }  \\ &\times & 
\langle \pi^+ (p') | \bar q (0,
-\frac{z^-}{2} , {\bf{b}} ) \gamma^+ q (0, \frac{z^-}{2} , {\bf{b}} )
| \pi^+ (p) \rangle , \nonumber 
\end{eqnarray} 
where $x$ is the Bjorken $x$, $\Delta_\perp=p'-p$ lies in the
transverse plane, and $b$ is an impact parameter. The model-independent
relation found in Ref.~\cite{Miller:2007uy} reads in the pion case
\begin{eqnarray}
\int_0^1 dx q(x,{\bf b}) = \int \frac{d^2 {\bf q}_\perp }{(2\pi)^2} e^{ i {\bf q}_\perp
\cdot {\bf b} } F_V (-{\bf q}_\perp^2)  \label{impact-ff}
\end{eqnarray} 

By crossing, the process related to the Deeply Virtual Compton Scattering (DVCS) 
off the pion, {\em i.e.}, two pion
production in $\gamma^* \gamma $ collisions, can be measured at low
invariant masses~\cite{Diehl:1998dk}. The relevant matrix element
reads
\begin{eqnarray}
&& \Phi^{ab} ( u, \zeta, W^2)  = \int
\frac{dz^-}{4\pi} e^{i x p^+ z^-} \times \label{defcrossedGPD} 
\\ && \;\;\;
\left . \langle \pi^a (p_1) \pi^b( p_2) | \bar \psi (0) \gamma \cdot n \,T\, \psi (z)
| 0 \rangle \right |_{z^+=0,z^\perp=0}, \nonumber 
\end{eqnarray} 
where $W^2 =(p_1+p_2)^2$, $\zeta= p_1 \cdot n / P \cdot n$ and
$u=(p_1-p_2)^2$.  By comparing, we have
\begin{eqnarray} 
\Phi^{ab} (u , \zeta, W^2 ) = {\cal H}^{ab} (x,\zeta,t) 
\end{eqnarray} 
One has the {\it soft pion theorem}~\cite{Polyakov:1998ze},
\begin{eqnarray} 
\Phi^{I=1} (u , 1, 0 ) = H^{I=1} (2u-1,1,0)= \phi (u),
\label{eq:soft}  
\end{eqnarray} 
where $\phi(u)$ represents the Pion Distribution Amplitude
(PDA) defined as 
\begin{eqnarray}
&&\langle \pi^a(p) | \bar \psi (z) \gamma^\mu \gamma_5 \frac12 \tau^b \psi(0)
| 0 \rangle  |_{z^+=0,z^\perp=0} \nonumber \\ &&= i f p^\mu \delta^{ab} \int_0^1 dx e^{i x p \cdot z}\phi(x).
\end{eqnarray}  
Note that result (\ref{eq:soft}) is based on PCAC and hence is a consequence of
the chiral symmetry.  One of the reasons to prefer GPD's rather than
2$\pi$PDA is the absence of final state interactions, which are
suppressed in the large $N_c$ limit\footnote{The simplest example
illustrating this feature is provided by the pion electromagnetic form factor. The
radius reads
$$ \langle r^2 \rangle_\pi = \frac{6}{M_V^2} \left[1 - \frac{1}{ 4
N_c} \log \left( \frac{m_\pi^2}{M_V^2}\right) \right],
$$ the first contribution stemming from the quark loop and the second
contribution an estimate from pion loops~\cite{RuizArriola:1991bq}.}.

\subsection{Formal results for chiral quark models}

\begin{figure}[tb]
\includegraphics[width=5.8cm]{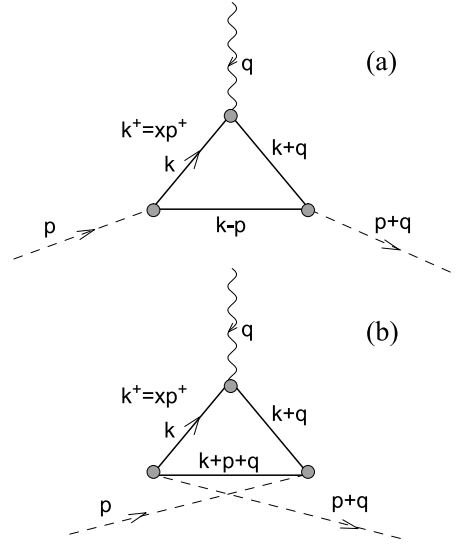} 
\caption{The direct (a) and crossed (b) Feynman diagrams for the
quark-model evaluation of the GPD of the pion.
\label{fig:diag}}
\end{figure}

\begin{figure}[tb]
\includegraphics[width=3.7cm]{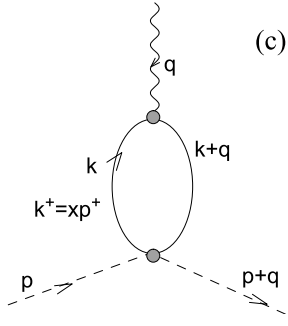} 
\caption{The contact contribution (c) to the GPD of the pion, responsible for the $D$-term. 
\label{fig:diag2}}
\end{figure}

The reduction formulas applied to the definition (\ref{defGPD01})
result in the amputated three-point Green function with the
constrained quark momentum integration, $k^+=x p^+$.  Large-$N_c$ treatment
leads to one-quark-loop diagrams, with massive quarks due to
spontaneous chiral symmetry breaking.  In {\em nonlinear} chiral quark
models the quark-pion interaction is described by the term $- \bar \psi
\omega U^5 \psi$ in the effective action, where the pion field matrix
is 
\begin{eqnarray}
U^5 = \exp( i \gamma_5 \tau \cdot \phi /f), \label{U5}
\end{eqnarray} 
where$f$ denotes the pion
decay constant.  The resulting Feynman rules and the definition
(\ref{defGPD}) lead to the Feynman diagrams of Figs.~\ref{fig:diag}
and \ref{fig:diag2}\footnote{The similar calculation of Ref.~\cite{Theussl:2002xp} uses the linear
realization of the chiral symmetry, with the $\sigma$ field
present.}. The presence of the contact term is crucial for the
preservation of the chiral symmetry \cite{Polyakov:1999gs}.  The evaluation of the diagrams
is straightforward, giving the following result for the isosinglet and
isovector parts:
\begin{eqnarray}
%{\cal H}^{ab}(x,\zeta,t)&=&\delta^{ab}{\cal H}_{I=0}(x,\zeta,t)+i \epsilon^{cab}\tau^c {\cal H}_{I=1}(x,\zeta,t), \nonumber\\
{\cal H}_{I=0}(x,\zeta,t)&=&{\cal H}_a(x,\zeta,t)+{\cal H}_b(x,\zeta,t)+{\cal H}_c(x,\zeta,t), \nonumber \\
{\cal H}_{I=1}(x,\zeta,t)&=&{\cal H}_a(x,\zeta,t)-{\cal H}_b(x,\zeta,t).  \label{defHabc}
\end{eqnarray}

The explicit contributions of the subsequent diagrams to the GPD's of
Eq.~(\ref{defHabc}) are
\begin{eqnarray}
&&\!\!\!\!\!\!\!{\cal H}_a(x,\zeta,t)=\frac{i N_c \omega^2}{4 \pi^2
f^2} \int d^4k \delta(k\cdot n -x) \times \nonumber \\ &&
\!\!\!\!\!\!\frac{\omega ^2\!-\!k^{2}\!-\!\zeta
\left(\omega^2\!-\!k^{2} \!+\!k\cdot p\right)\!+\!x
\left(\omega^2\!-\!\frac{t}{2}\!-\!k^{2}\!+\!2 k\cdot p\right)
\!-\!k\cdot q}{ D_k D_{k+q} D_{k-p}}, \nonumber \\
&&\!\!\!\!\!\!\!{\cal H}_b(x,\zeta,t)=\frac{i N_c \omega^2}{4 \pi^2}
f^2 \int d^4k \delta(k\cdot n -x) \times \nonumber \\
&&\!\!\!\!\!\!\frac{\!-\!\omega ^2\!+\!k^{2}\!+\!\zeta k \cdot p\!+\!x
\left(\omega^2\!-\!  \frac{t}{2}\!-\!k^{2}\!-\!2 k\cdot p\!-\!2 k\cdot
q\right)\!+\!k\cdot q}{D_k D_{k+q} D_{k+p+q}}, \nonumber \\
&&\!\!\!\!\!\!\!{\cal H}_c(x,\zeta,t)=\frac{i N_c \omega^2}{4\pi ^2
f^2} \int d^4k\delta(k\cdot n -x) \frac{2 x-\zeta}{D_k D_{k+q}}.
\end{eqnarray}
The denominator of the proparator of quark of mass $\omega$ and
momentum $l$ is denoted as
\begin{eqnarray}
{D_l}={l^2-\omega^2+i0}. \label{defD}
\end{eqnarray}
The powers of the momentum $k$ in the numerators may be eliminated
with the following reduction formulas:
\begin{eqnarray}
k^2&=&D_k+\omega^2, \nonumber \\ k\cdot q&=&\frac{1}{2}(D_{k+q}-t-D_k), \nonumber \\ 
k\cdot p&=&-\frac{1}{2}(D_{k-p}-m_\pi^2-D_k), \nonumber \\
k\cdot p&=&\frac{1}{2}(D_{k+p+q} - m_\pi^2 - D_{k+q} + t). \label{tricks}
\end{eqnarray}
Then the GPD's become  
\begin{eqnarray}
&&\!\!\!\!\!\!\! {\cal H}_{I=0,1}(x,\zeta,t)=\frac{-i N_c \omega^2}{8
\pi^2 f^2} \int d^4k \delta(k\cdot n -x) \times \label{H01} \\
&&\!\!\!\!\!\!\! \left ( \frac{1}{D_k D_{k-p}}
+\frac{1-\zeta}{D_{k+q}D_{k-p}} \mp \frac{1}{D_{k+q} D_{k+p+q}} \mp
\frac{1-\zeta}{D_{k} D_{k+p+q}} \right . \nonumber \\ &&\!\!\!\!\!\!\!
\left . + \frac{(\zeta -2 x) m_\pi^2+t (x-1)}{ D_k D_{k+q} D_{k-p}}
\mp \frac{(\zeta -2 x) m_\pi^2+t (x-\zeta +1)}{ D_k D_{k+q} D_{k+p+q}}
\right ), \nonumber
\end{eqnarray}
with the upper (lower) signs corresponding to the case of $I=0$
($I=1$).  Note that the piece with $1/(D_k D_{k+q})$ cancels out due
to the presence of contact diagram (c). The contribution of the
diagram (c), having the support for $x \in [0,\zeta]$, is the
$D$-term \cite{Polyakov:1999gs}.

\begin{figure}[tb]
\begin{center}
\includegraphics[width=5.3cm]{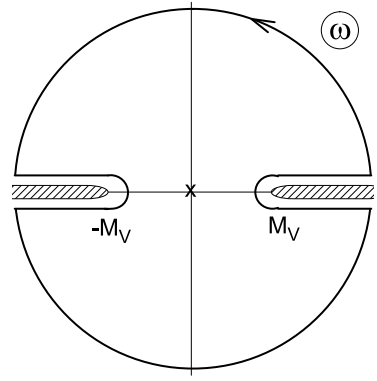}
\end{center}
\caption{The contour $C$ for evaluation of observables in the meson
dominance variant of SQM. $M_V$ denotes the $\rho$-meson mass. The
cross and hatched regions indicate the position of the pole and cuts
of the spectral function Eq.~(\ref{rhov}).
\label{contour0}}
\end{figure}

From the above form it is clear that we need to consider two generic
types of two- and three-point integrals:
\begin{eqnarray}
&&I(x,l\cdot n,l'\cdot n,(l-l')^2)=\frac{-i N_c \omega^2}{4 \pi^2 f^2}
\int d^4k \frac{\delta(k\cdot n -x)}{D_{k-l} D_{k-l'}}, \nonumber \\
&&J(x,l\cdot n,l'\cdot n,l^2,l'^2,l \cdot l')=\nonumber \\ &&
\;\;\;\;\;\;\;\; \frac{i N_c \omega^2}{4 \pi^2 f^2} \int d^4k
\frac{\delta(k\cdot n -x)}{D_k D_{k-l} D_{k-l'}}. \label{defIJ}
\end{eqnarray} 
These are analyzed in detail Appendices~\ref{sec:2p} and \ref{sec:3p}.
The two-point function $I$ is logarithmically divergent, hence the
analysis needs regularization. This is where different quark models
depart from one another. We may separate the issues of regularization
from formal expressions, which is convenient for theoretical aspects
and the demonstration of the consistency conditions.  Written in terms
of the basic two- and three-point functions Eq.~(\ref{H01}) become
\begin{eqnarray}
&&\!\!\!\!\!\!\! {\cal H}_{I=0,1}(x,\zeta,t)= \label{H02} \frac{1}{2} \left [ I(x,0,1,m_\pi^2) + (1-\zeta) I(x,\zeta,1,m_\pi^2) 
\right . \nonumber \\ && \!\!\!\!\! \left . \mp I(x,-1+\zeta,\zeta,m_\pi^2) 
\mp (1-\zeta)I(x,-1+\zeta,0,m_\pi^2)
\right . \nonumber \\ 
&&\!\!\!\!\!\!\! \left . - [(\zeta -2 x) m_\pi^2+t (x-1)]J[x,\zeta,1,t,m_\pi^2,-\frac{t}{2}] \right . \label{generic} \\ && \!\!\!\!\!\left. 
\pm [(\zeta -2 x) m_\pi^2+t (x-\zeta +1)]J[\zeta-x,\zeta,1,t,m_\pi^2,-\frac{t}{2}] 
\right ]. \nonumber
\end{eqnarray}
This equation may be considered as the generic non-linear local chiral
quark-model result for the isospin-projected GPD's of the pion. Model
details, such as regularization, affect the specific form of the two-
and three-point functions, but leave the structure of
Eq.~(\ref{generic}) unchanged. The nontrivial features of the
regularization will utterly be responsible for the fulfillment of the
general properties of GPD's described in Section~\ref{sec:formalism}. 

\section{Results of the Spectral Quark Model \label{sec:sqm}}

\begin{figure}[tb]
\includegraphics[width=7.7cm]{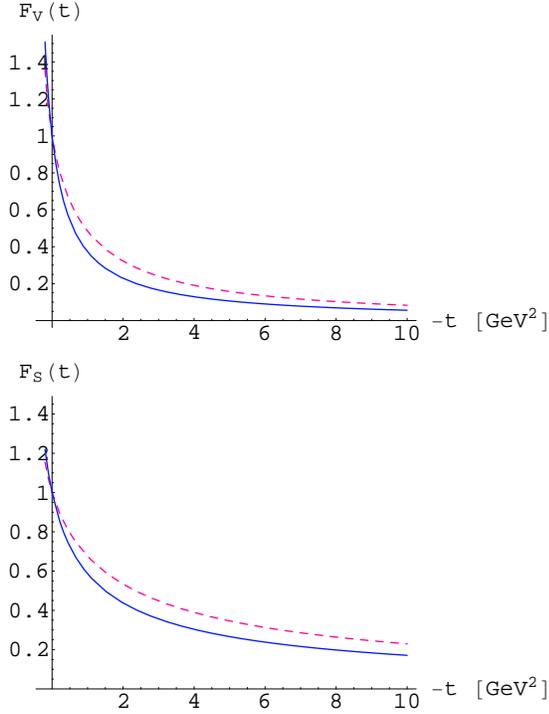} \vspace{-7mm}
\caption{(Color online) The pion electromagnetic, $F_V(t)$, (top) and gravitational,
$\theta_1(t)=\theta_2(t)\equiv F_S(t)$, (bottom) form factors in SQM (solid line,
Eq.~(\ref{ffSQM},\ref{ffSQMg})) and in NJL model (dashed line,
Eq.~(\ref{ffnjl})). \label{fig:pff}}
\end{figure}

Now we come to the evaluation of GPD in specific models. From now on
we work for simplicity in the chiral limit,
\begin{eqnarray}
m_\pi=0.
\end{eqnarray} 
The first model we consider is the Spectral Quark Model (SQM) of
Ref.~\cite{RuizArriola:2003bs}, where all the necessary details of the
model can be found. The one-quark-loop action of this model is
\begin{eqnarray}
\Gamma =-i N_c \int_C d \omega \rho(\omega) {\rm Tr} \log
\left( i\slashchar{\partial} - \omega U^5 \right),
\label{eq:eff_ac} 
\end{eqnarray} 
where $\rho(\omega)$ is the quark generalized {\em spectral
function}, and $U^5$ is given in Eq.~(\ref{U5}). 
In the calculations of this paper we only need the vector
part of the spectral function, which in the {\em meson-dominance} SQM
\cite{RuizArriola:2003bs} has the form
\begin{eqnarray}
\rho_V (\omega) &=& \frac{1}{2\pi i} \frac{1}{\omega}
\frac{1}{(1-4\omega^2/M_V^2)^{5/2}},  \label{rhov}
\end{eqnarray}
exhibiting the pole at the origin and cuts starting at $\pm M_V/2$,
where $M_V$ is the mass of the vector meson, $M_V=m_\rho=770$~MeV. The
contour $C$ for the integration in (\ref{eq:eff_ac}) is shown in
Fig.~\ref{contour0}. 
Despite the rather unusual appearance of the spectral function, the model leads to 
conventional phenomenology \cite{RuizArriola:2003bs,Arriola:2006ds}. Importantly, it implements the vector-meson dominance,
yielding the pion electromagnetic form factor of the monopole form
\begin{eqnarray}
F_V^{\rm SQM}(t)=\frac{M_V^2}{M_V^2-t}. \label{ffSQM}
\end{eqnarray}
For the gravitational form factor we find
\begin{eqnarray}
\theta_1^{\rm SQM}(t)=\theta^{\rm SQM}_2(t)=\frac{M_V^2}{t} \log \left
( \frac{M_V^2}{M_V^2-t} \right ) \equiv F_S^{\rm SQM}(t). \nonumber \\ \label{ffSQMg}
\end{eqnarray}
Both the electromagnetic and gravitational form factors for SQM are plotted in Fig.~\ref{fig:pff} with solid lines.

With the help of Eq.~(\ref{s1},\ref{chi}) it is straightforward to
obtain the formulas for the GPD's in SQM. The expressions are simple in
the chiral limit, and shortest in the asymmetric notation. For the
quark and antiquark GPD's we obtain
\begin{widetext}
\begin{eqnarray}
&& 2{\cal H}_q(x,\zeta,t)=\theta ((1-x) x)+\theta ((1-x) (x-\zeta ))+t
   (1-x) \left [ \frac{2 (x-1) \left(t (x-1)^2+3 (\zeta -1)
   M_V^2\right) \theta (1-x) \theta (x-\zeta )}{\left(t (x-1)^2+(\zeta
   -1) M_V^2\right)^2} \right .  \nonumber \\ && \left . +
   \left(\frac{(x-1) \left(t (x-1)^2+3 (\zeta -1)
   M_V^2\right)}{\left(t (x-1)^2+(\zeta -1) M_V^2\right)^2}+\frac{(x
   (\zeta -2)+\zeta ) \left(3 (\zeta -1) \zeta ^2 M_V^2+t
   \left(\left(\zeta ^2+8 \zeta -8\right) x^2+2 (4-5 \zeta ) \zeta
   x+\zeta ^2\right)\right)}{\left(\zeta ^2+\frac{4 t x (x-\zeta
   )}{M_V^2}\right)^{3/2} \left(t (x-1)^2+(\zeta -1)
   M_V^2\right)^2}\right) \right . \nonumber \\ &&
   \;\;\;\;\;\;\;\times \left .  \theta (x) \theta (\zeta -x)\right],
   \nonumber \\ && 
   \nonumber \\ && {\cal H}_{\bar q}(x,\zeta,t)={\cal
   H}_{q}(\zeta-x,\zeta,t). \label{HHb}
\end{eqnarray}
\end{widetext}
From these, the isospin combinations are trivial to get.  The formulas satisfy the
consistency relations (\ref{eq:sym},\ref{norm},\ref{norm2}). In
particular, upon passing to the symmetric notation and using the above
formulas we verify
\begin{eqnarray}
&&\int_{-1}^1 dX\, {H}^{I=1}(X,\xi,t)= 2F_V^{\rm SQM}(t), \nonumber \\
&&\int_{-1}^0 dX\, X\,{H}^{I=0}(X,\xi,t)=(1-\xi^2) F_S^{\rm SQM}(t). \label{sumsqm}
\end{eqnarray}
The isovector norm is decomposed as follows between the ERBL and DGLAP
regions:
\begin{eqnarray}
&&\!\!\!\!\!\!\!\!\int_0^\zeta \!\!dx\, {\cal H}_q(x,\zeta,t)=\frac{2-\zeta}{2} \frac{M_V^2}{M_V^2-t} 
\frac{\zeta \left( M_V^2 + t (1-\zeta)\right)}{(2-\zeta) \left(M_V^2-t (1-\zeta)\right)}, \nonumber \\
&&\!\!\!\!\!\!\!\!\int_\zeta^1 \!\!dx\, {\cal H}_q(x,\zeta,t)=\frac{2-\zeta}{2} \frac{M_V^2}{M_V^2-t} 
\frac{2 (1-\zeta) \left(M_V^2-t\right)}{(2-\zeta) \left(M_V^2-t (1-\zeta)\right)}. \nonumber \\
&&\!\!\!\!\!\!\!\!\!~
\end{eqnarray}

Some special values of the GPD's in SQM are:
\begin{eqnarray}
&& {\cal H}^{I=1}(1,\zeta,t)={\cal H}^{I=0}(1,\xi,t)=1, \label{spec}
\\ && {\cal H}^{I=1}(\zeta,\zeta,t)=\frac{M_V^2
\left(M_V^2+t(1-\zeta)\right)}{\left(M_V^2-t (1-\zeta)\right)^2},
\nonumber \\ && \lim_{x \to \zeta^+}{\cal
H}^{I=0}(x,\zeta,t)=\frac{M_V^2
\left(M_V^2+t(1-\zeta)\right)}{\left(M_V^2-t (1-\zeta)\right)^2},
\nonumber \\ && \lim_{x \to \zeta^-}{\cal H}^{I=0}(x,\zeta,t)=\frac{t
\left(3 M_V^2-t(1-\zeta)\right) (1-\zeta)}{\left(M_V^2-t
(1-\zeta)\right)^2}. \nonumber
\end{eqnarray}
The values at $x=-1+\zeta$ and $x=0$ follow from the symmetry
relations ${\cal H}^{I=1,0}(x,\zeta,t)=\pm {\cal
H}^{I=1,0}(\zeta-x,\zeta,t)$.  We note that the discontinuities at the
end points $x=\pm 1$ and for the $I=0$ part at $x=0$ and $x=\zeta$ are
a typical feature of quark-model calculations. The QCD evolution
immediately washes out these discontinuities, see
Sect.~\ref{sec:evol}.  The derivative of ${\cal H}^{I=1}(x,\zeta,t)$
with respect to $x$ is continuous at the point $x=\zeta$, where
\begin{eqnarray}
&& \frac{d}{dx}{\cal H}^{I=1}(x,\zeta,t)\mid_{x=\zeta}=-\frac{2 M_V^2
t \left(3 M_V^2+t(1-\zeta)\right)}{\left(M_V^2-t (1-\zeta)\right)^3}.
\nonumber \\
\end{eqnarray}

On general grounds, Eqs.~(\ref{HHb}) also satisfy the polynomiality
conditions (\ref{poly}), which can be seen from the derivation through
the double distributions shown in Appendix \ref{sec:alpha}.  Also note
that the obtained formulas are certainly not of the form where the
$t$-dependence is factorized, {\em i.e.}
\begin{eqnarray}
H^{I=0,1}(X,\xi,t)\neq F(t) G(X,\xi).
\end{eqnarray}

For the case of $t=0$ the formulas (\ref{HHb}) simplify to the
well-know \cite{Polyakov:1999gs,Theussl:2002xp} step-function results
\begin{eqnarray}
&&\!\!\!\!\!\!\! {\cal H}_{I=0}(x,\zeta,0) = \theta[(1 - x)(x - \zeta )] - 
\theta[-x(x + 1 - \zeta )], \label{t0} \nonumber \\
&&\!\!\!\!\!\!\! {\cal H}_{I=1}(x,\zeta,0)=\theta[(1-x)(x+1-\zeta)].  \label{zerot}
\end{eqnarray}

Another simple case is for $\zeta=0$ and any value of $t$,
\begin{eqnarray}
{\cal H}_{q}(x,0,t) =\frac{M_V^2 \left(M_V^2+t
(x-1)^2\right)}{\left(M_V^2-t (x-1)^2\right)^2},
\end{eqnarray}
which agrees with the result reported in \cite{Broniowski:2003rp}. The
corresponding impact parameter representation obtained there is given
by the formula\footnote{Note an overall sign missing in
Ref.~\cite{Broniowski:2003rp}.}
\begin{eqnarray}
q(x,{\bf b}) &=&  \frac{M_V^2}{2\pi(1-x)^2} \times \label{eq:impact0} \\ && \left [ 
\frac{b M_V}{1-x} K_1 \left ( \frac{b M_V}{1-x} \right) - K_0 \left ( \frac{b M_V}{1-x} \right) \right ]. \nonumber 
\end{eqnarray} 
From this expression one obtains
\begin{eqnarray}
\int_0^1 dx\, q(x,{\bf b}) =  \frac{M_V^2 K_0 (b M_V)}{2\pi}. \label{eq:impact}
\end{eqnarray} 
This complies to the model-independent
relation~(\ref{impact-ff}) when the vector-dominance form factor (\ref{ffSQM}) 
is used, since, explicitly, 
\begin{eqnarray}
\int \frac{d^2 q }{2\pi} \frac{e^{ i {\bf q}\cdot {\bf b} } }{M_V^2+{\bf q}^2}= K_0 (b M_V).
\end{eqnarray}

\begin{figure}[tb]
\includegraphics[width=8.5cm]{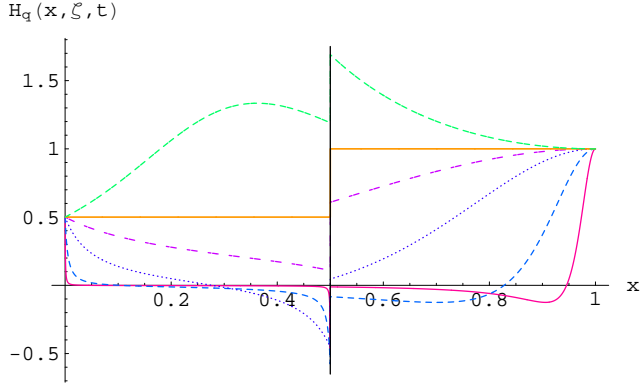} 
\caption{(Color online) The SQM results for the quark GPD of the pion, ${\cal H}_q$ of Eq.~(\ref{defH}),
plotted as a function of $x$ for $\zeta=1/2$ and \mbox{$t=0.2,0,-0.2,-1,-10,-100~{\rm GeV}^2$}, from top
to bottom (at $x=0.9$). Asymmetric notation. \label{fig:Hq}}
\end{figure}

\begin{figure}[tb]
\includegraphics[width=8.9cm]{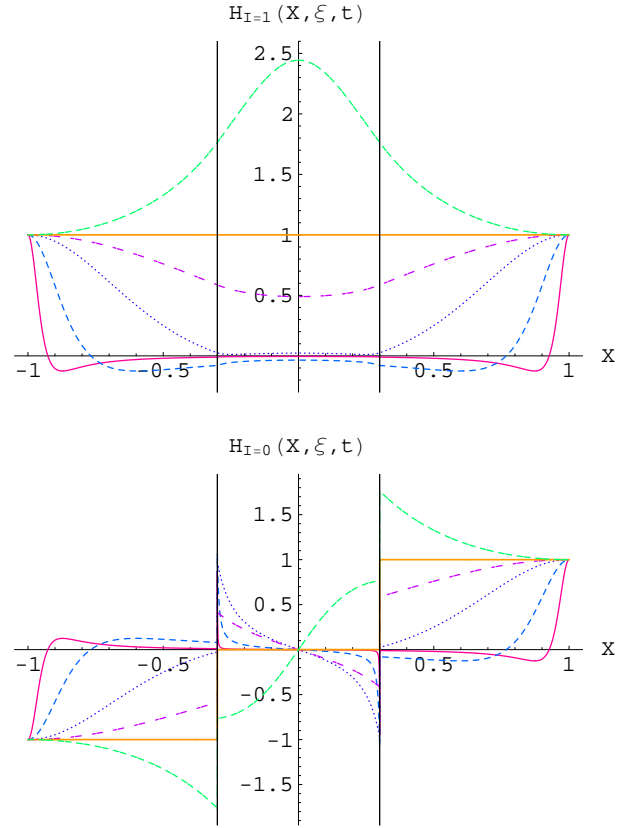}
\caption{(Color online) Same as Fig.~\ref{fig:Hq} for ${H}_{I=1}$ (top) and
${H}_{I=0}$ (bottom) in the symmetric notation,
$\xi=1/3$. \label{fig:H01}}
\end{figure}

The case of Eq.~(\ref{HHb}) for $\zeta=0.5$ and several values of $t$
is shown in Fig.~\ref{fig:Hq}.  Results for other values of $\zeta$
are qualitatively similar. Figure~\ref{fig:H01} shows the isospin 0
and 1 combinations, ${H}_{I=0,1}$. We note that the $I=1$ GPD and its
first derivative with respect to $x$ is continuous at $x=0$ and
$x=\zeta$, while the $I=0$ combination is discontinuous at these
points.

At $t=0$ we have the above-mentioned 
step-function results 
\begin{eqnarray}
&&{H}_{I=1}=\theta(1-X^2), \\
&&{H}_{I=0}=\theta((1-X)(X-\xi))- \theta((1+X)(-X-\xi)).\nonumber 
\end{eqnarray} 
As $-t$ increases, the 
strength moves to the vicinity of the $X=\pm 1$ points. 
The limit of $-t(1-x) \to \infty$ in Eqs.~(\ref{HHb}) yield the asymptotic forms 
\begin{eqnarray}
&&{\cal H}^{I=1,0}(1,\zeta,t)=1, \label{asyt} \\
&&{\cal H}^{I=1}(x,\zeta,t)\simeq \frac{M_V^2(1-\zeta)}{t(1-x)^2}, \;\;\; x\in[0,1), \nonumber \\
&&{\cal H}^{I=0}(x,\zeta,t)\simeq \frac{M_V^2(1-\zeta)}{t(1-x)^2}, \;\;\; x\in(\zeta,1), \nonumber \\
&&\lim_{x\to \zeta^-}{\cal H}^{I=0}(x,\zeta,t)=-1+\frac{M_V^2}{(1-\zeta) t} , \nonumber \\
&&{\cal H}^{I=0}(x,\zeta,t)\simeq \frac{M_V^2 (2 x-\zeta) \left(\zeta^2-3 \zeta+2\right)}
{2 (x-1)^2 (x-\zeta+1)^2 t}, \;\;\; x\in(0,\zeta). \nonumber
\end{eqnarray}
In the DGLAP region the absolute value of the $I=0,1$ functions are
bounded by unity. Note that at large $-t$ the GPD's continue
to be equal to 1 at $x=1$, however very quickly drop to zero in the
DGLAP region. In the ERBL region the $I=1$ part drops, while the $I=0$
part tends to $-1$ as $x \to \zeta^-$, and drops to 0 elsewhere.

Since in our quark model the parton distribution function is unity,
$q(x)=1$, the positivity bound~(\ref{positiv}) states that (for $t \le 0$)
\begin{eqnarray}
|H^{\rm SQM}_q(X,\xi,t)| \le 1, \;\;\;\;\; \xi \le X \le 1. \label{positiv2} 
\end{eqnarray} 
It is a priori not obvious that the bound should hold in chiral quark
models where finiteness of observables results from
regularization involving subtractions. Nevertheless, we have checked
with Eqs.~(\ref{HHb}) that condition (\ref{positiv2}) is actually
satisfied in the DGLAP region for all values of $\xi$ and all negative
$t$. This is also manifest in Fig.~\ref{fig:Hq}, as well as in
Eq.~(\ref{asyt}). The bound is saturated at the end points $X=\pm
1$. Thus the positivity bound is satisfied in SQM.

\begin{figure}[tb]
\includegraphics[width=8.9cm]{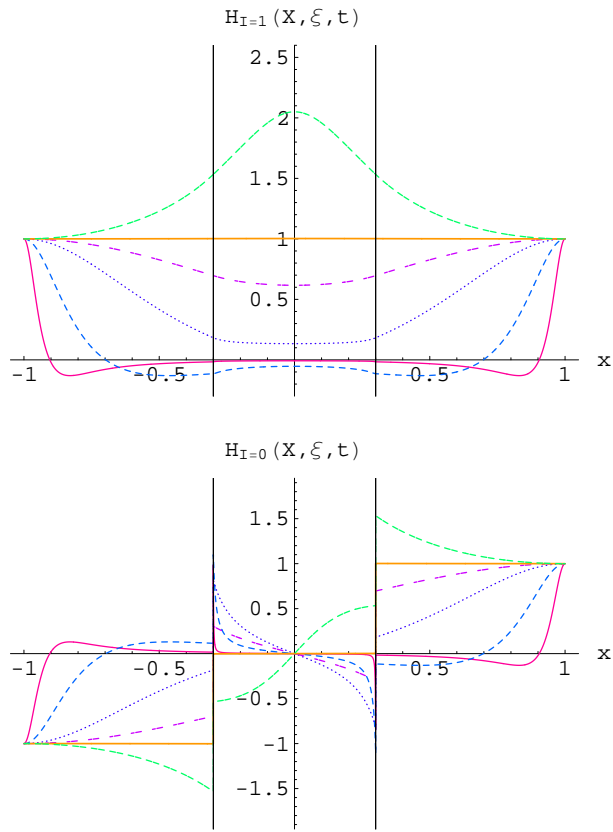} 
\caption{(Color online) Same as Fig.~\ref{fig:H01} for for the NJL model with the
PV regularization, $M=280$~MeV, $\Lambda=871$~MeV.
\label{fig:H01NJL}}
\end{figure}

\section{Results of the Nambu--Jona-Lasinio model \label{sec:njl}}

We use the non-linear NJL model with Pauli-Villars (PV) regularization in the twice-subtracted version of 
Ref.~\cite{RuizArriola:2002wr}. The prescription for regularizing an observable ${\cal O}$ in this model is 
\begin{eqnarray}
{\cal O}_{\rm reg} = {\cal O}(0) - {\cal O}(\Lambda^2 ) + \Lambda^2
\frac{d {\cal O}(\Lambda^2 )}{d\Lambda^2}, \label{prescr}
\label{eq:PV2} 
\end{eqnarray} 
where $\Lambda$ are the PV regulator. Note that Eq.~(\ref{prescr}) is
different from the prescription used in \cite{Theussl:2002xp}, where a
variant of the PV regularization with two distinct cut-offs
is applied. We also use the non-linear rather than linear realization
of the chiral field.  In what follows we take $M=280$~MeV for the
quark mass and $\Lambda=871$~MeV, which yields $f=93.3$~MeV
\cite{RuizArriola:2002wr} according the the formula
\begin{eqnarray}
f^2=-\frac{3M^2}{4\pi^2} \left ( \log(\Lambda^2 + M^2) \right )_{\rm reg}. \label{f2njl}
\end{eqnarray} 

The pion electromagnetic form factor in the NJL model is
\begin{eqnarray}
&&F_V^{\rm NJL}(t)=1+ \frac{N_c M^2}{8 \pi^2 f^2} \times \label{ffnjl} \\
&&\left ( 
\frac{2 \sqrt{4 \left(M^2+\Lambda ^2\right)-t} \log \left(\frac{\sqrt{4 \left(M^2+\Lambda ^2\right)-t}-\sqrt{-t}}{\sqrt{4
   \left(M^2+\Lambda ^2\right)-t}+\sqrt{-t}}\right)}{\sqrt{-t}}
\right )_{\rm reg}, \nonumber
\end{eqnarray}
The property $\lim_{t \to - \infty} F_{\rm NJL}(t)=0$ follows from
Eq.~(\ref{f2njl}).  The isovector form factors arising in both
considered models are compared in Fig.~\ref{fig:pff}.  The formula for
the gravitational form factor in the NJL is lengthy, hence we only give
the numerical results in Fig.~\ref{fig:pff}. In this model also the
two gravitational form factors are equal to each other, $\theta_1^{\rm
NJL}(t)=\theta_2^{\rm NJL}(t)$. We note that although the form factors
in both models are qualitatively similar, they are quantitatively
somewhat different, which is partly due to the choice of parameters in
the NJL model, as well as follows from different analytic structure of
the corresponding formulas, in particular at large values of $-t$.

The application of the formulas derived in Appendix~\ref{app:njl}
leads to expressions similar to those of \cite{Theussl:2002xp}.
The analogs of Eq.~(\ref{spec},\ref{asyt}) in the NJL model are more
complicated, hence we do not give them here.  More details may be
found in Ref.~\cite{Theussl:2002xp}.  For the special case of $t=0$
Eq.~(\ref{t0}) holds.  As in SQM, the conditions
(\ref{norm},\ref{norm2},\ref{eq:sym}) are satisfied in the considered
NJL model with the PV regularization.

The numerical results for the NJL model are displayed in
Fig.~\ref{fig:H01NJL}.  When comparing Figs.~\ref{fig:H01} and
\ref{fig:H01NJL} we note a striking similarity between the two
considered quark models.  The slight differences stem mainly from
different form factors in the two considered models, {\em cf.}
Fig.~\ref{fig:pff}.  Our results are also qualitatively similar to the
case of the chiral limit in Fig.~6 of Ref.~\cite{Theussl:2002xp}.  As pointed
out in Appendix~\ref{app:njl}, polynomiality follows from the
derivation proceeding via the double distributions.  We have checked
that the similarly to SQM, the positivity bound (\ref{positiv}) is
also satisfied in the NJL model at any $\xi$ and all negative values
of $t$.

\section{QCD evolution of quark models}
\label{sec:qcd-qm}

\subsection{The need for evolution}

A key question, not only for our model but for any non-perturbative
calculation, is {\it what} is the scale at which our model result for
the GPD's holds. Ultimately, this boils down to the issue on {\it how}
the model predictions for the GPD's might be confronted to
experimental or lattice data. In QCD, the GPD's are scale dependent, while
in models they correspond to functions defined at a given scale. This
is so because low energy models hold at a scale above which scaling
should set in. Perturbative QCD and the corresponding scaling
violations bring in the issue of evolution equations for GPD's which
will be treated in detail in Sec.~\ref{sec:evol}. In this section we
discuss and update the procedure already used in previous
works~\cite{Davidson:1994uv,Davidson:2001cc,RuizArriola:2002bp,
RuizArriola:2002wr} for the evolution of PDF and PDA and extract its
consequences as compared to available experimental data or lattice
results.

From the point of view of perturbative QCD where both quarks and
gluons contribute as {\it explicit} degrees of freedom, the role of
the low-energy chiral quark models is to provide initial conditions
for the QCD evolution equations order by order in the twist expansion.
Clearly, chiral quark models contain non-perturbative QCD features,
particularly the spontaneous chiral symmetry breaking. On the other
hand, chiral quark models do not contain the QCD degrees of freedom,
{\em i.e.}, the current quarks and explicit gluons. So one expects
typical high-energy perturbative QCD features, such as radiative
corrections, to be absent in the model. This is precisely the pattern
of logarithmic scaling violations which the models lack but which have
traditionally been computed in the perturbation theory in QCD.

The procedure applied in this paper takes the quark-model
distributions at some low quark-model scale $Q_0$ and evolves them to
higher scales, where (for some observables) the experimental or
lattice data are available.  In the following we use the leading-order
ERBL-DGLAP evolution equations with three flavors.  This strategy
reflects the present state of the art, which can be validated by
comparing our predictions both to experiment as well as available
lattice data. It should be noted, however,
that to date there is no rigorous relation between the QCD quarks and
the constituent quarks of the chiral models, and a more fundamental
description of the transition from the hard to the soft would be very
helpful.

\subsection{Momentum fraction and the matching condition}

For definiteness, we consider $\pi^+$, and denote $q(x)$ and $\bar q(x)$ 
the single-flavor distributions of quarks and antiquarks. 
The valence (or non-singlet) quark distribution is 
\begin{eqnarray}
V=u_\pi - \bar u_\pi + \bar d_\pi - d_\pi, \label{val}
\end{eqnarray}
while the non-singlet quark distribution is 
\begin{eqnarray}
S=u_\pi + \bar u_\pi + d_\pi +  \bar d_\pi + s_\pi + \bar s_\pi . 
\end{eqnarray}
The sea quark distribution is defined as 
\begin{eqnarray}
s=S-V=2( \bar u_\pi + d_\pi) + s_\pi + \bar s_\pi .
\end{eqnarray}
Isospin and crossing symmetries implies the property
\begin{eqnarray}
u_{\pi^+} (x) = \bar d_{\pi^+} (1-x). \label{crossing}
\end{eqnarray}

The energy-momentum tensor $\Theta^{\alpha \beta}$ is a conserved
quantity in any relativistic theory and hence renormalization
invariant, due to the Poincar\'e invariance. Its diagonal matrix element
between the pion state of momentum $p$ is
\begin{eqnarray}
\langle \pi(p) \, | \Theta^{\alpha \beta} | \pi (p) \,  \rangle = 2 p^\alpha p^\beta.  
\end{eqnarray} 
For the QCD Lagrangian the energy-momentum tensor can be separated
into several contributions in a gauge-invariant but scale- and hence
scheme-dependent manner~\cite{Ji:1995sv}. Although we will be
considering the LO evolution, for our purposes we may have the
standard $\overline{\rm MS}$ scheme in mind and write
\begin{eqnarray}
\Theta^{\alpha \beta}= 
\Theta_g^{\alpha \beta} + \Theta_s^{\alpha \beta} + \Theta_v^{\alpha \beta}
\end{eqnarray} 
where $\Theta_g^{\alpha \beta}$, $\Theta_s$, and
$\Theta_v^{\alpha \beta}$ are the gluon, sea-quark, and valence-quark contributions, respectively.
They are equal to
\begin{eqnarray}
\langle \pi | \Theta^{\alpha \beta}_g | \pi \rangle |_\mu &=& 2
p^\alpha p^\beta \langle x \, \rangle_g (\mu) \nonumber  \\
\langle \pi | \Theta^{\alpha \beta}_s | \pi \rangle |_\mu &=& 2
p^\alpha p^\beta \langle x \, \rangle_s (\mu) \nonumber  \\
\langle \pi | \Theta^{\alpha \beta}_v | \pi \rangle |_\mu &=& 2
p^\alpha p^\beta \langle x \, \rangle_v (\mu) \nonumber  \\
\end{eqnarray} 
where $ \langle x \, \rangle_g (\mu) $, $ \langle x \, \rangle_s (\mu)
$ and $ \langle x \, \rangle_v (\mu) $, are the gluon, sea quark, and
valence quark {\em momentum fractions} of the pion at the scale $\mu$,
respectively. In deep inelastic scattering (DIS) it can be
shown~\cite{Altarelli:1977zs} that if $q(x,\mu)$, $\bar q(x,\mu)$
and $ G(x,\mu)$ represent the probability density of finding a quark,
antiquark, and gluon, respectively, with the momentum fraction $x$ at the
scale $\mu$ (typically, we identify $\mu^2$ with $Q^2$ in DIS), then 
\begin{eqnarray}
\langle x \, \rangle_g (\mu) &=& \int_0^1 dx \, x \, G(x,\mu)  \\
\langle x \, \rangle_s (\mu) &=& \int_0^1 dx \, x \, s(x,\mu)  \\
\langle x \, \rangle_v (\mu) &=& \int_0^1 dx \, x \, V(x,\mu).  
\end{eqnarray} 
where, due to the crossing symmetry (\ref{crossing})
for a single flavor one has $\langle x \rangle_q = \langle x \rangle_u
= \langle x \rangle_{\bar d} = \langle x \rangle _v /2 $. 
The scale-dependent momentum fractions fulfill the
momentum sum rule
\begin{eqnarray}
\langle x \, \rangle_g (\mu) + 
\langle x \, \rangle_s (\mu) + 
\langle x \, \rangle_v (\mu)  = 1, \label{sr}  
\end{eqnarray} 
which is a consequence of the energy-momentum tensor conservation. In
perturbation theory due to radiative corrections
$\langle x \, \rangle_g (\mu) $ and $\langle x \, \rangle_s (\mu) $
decrease as the scale $\mu$ goes down. On the contrary, the
valence contribution to the energy momentum tensor evolves as
\begin{eqnarray}
\frac{ \langle x \, \rangle_v (Q) } { \langle x \, \rangle_v (Q_0)  } = \left( \frac{\alpha(Q)}
{\alpha(Q_0) } \right)^{\gamma_1^{\rm (0)} / (2 \beta_0) } \quad ,
\qquad
\end{eqnarray} 
where $ \gamma_1^{\rm (0)} / (2 \beta_0) = 32/81 $ for $N_F=N_c=3$.  
Downward LO QCD evolution would yield that for some given reference scale, $\mu_0 \equiv Q_0$,
\begin{eqnarray}
\langle x \, \rangle_v \, (Q_0) = 1 \qquad 
\langle x \, \rangle_s \, (Q_0) + \langle x \, \rangle_g \, (Q_0) = 0.
\end{eqnarray} 
The scale $Q_0$ defined with the above condition is called the {\it
quark model scale} for obvious reasons, as only valence quarks
contribute.  This may represent the matching condition between QCD and
the chiral quark models, schematically written as Eq.~(\ref{match}).

There exists a wealth of information on the momentum fraction carried
by valence quarks in the pion at scales $Q \sim 2~{\rm GeV}$, coming from
several sources. Phenomenological analyses require these high scales
to neglect higher twist corrections. The Durham
group~\cite{Sutton:1991ay}, based mainly on the E615 Drell-Yan
data~\cite{Conway:1989fs} and the model assumption that sea quarks
carry $10-20\%$ of the momentum fraction, determines $\langle x
\rangle_q= 0.235 (10)$ at the scale $Q= 2~{\rm GeV}$. The analysis
of Ref.~\cite{Gluck:1999xe}, based on the assumption that the momentum
fraction carried by valence quarks in the pion coindides with that of
the nucleon, yields $\langle x \rangle_q = 0.2$ at $Q= 2~{\rm GeV}$.

Other determinations, comprising lattice calculations, may access
directly the leading-twist contribution in a non-perturbative
manner. However, the transition from the intrinsically
non-perturbative lattice regularization to the perturbative
$\overline{\rm MS}$ regularization scheme requires high-energy
matching scales.  Early Euclidean lattice simulations provided
$\langle x \rangle_q = 0.32 (5)$ at the scale $Q^2 \approx 4.84 \pm
2.2 ~{\rm GeV}^2$~\cite{Martinelli:1987bh}. More recently, lattice
calculations linearly extrapolated to the chiral limit
~\cite{Best:1997qp}, yielding $ \langle x \rangle_q = 0.28 (1) $ at
the scale $Q^2 \approx 5.8~{\rm GeV^2}$, a somewhat larger value than
suggested by phenomenology \cite{Sutton:1991ay,Gluck:1999xe} and
expected from the quenched approximation. Still, in the quenched
approximation, in Ref.~\cite{Capitani:2005jp} $\langle x \rangle_q =
0.243(21)$ at $Q=2~{\rm GeV}$ for light pions, which is in a closer
agreement to the Durham~\cite{Sutton:1991ay} than to the
Dortmund~\cite{Gluck:1999xe} results. This value squares with the
gluon content of the pion $\langle x \rangle_g = 0.37 \pm 0.08_{\rm
stat} \pm 0.12_{\rm sys}$ at a similar scale but with $m_\pi \sim 900
~{\rm MeV}$, as extracted recently in Ref.~\cite{Meyer:2007tm}.

Finally, there exist transverse lattice calculations, where full
$x$-dependent parton properties can be determined non-perturbatively
at low scales~\cite{Burkardt:2001jg}. The calculation of
Ref.~\cite{Dalley:2001gj} gives $ \langle x \rangle_q = 0.43(1)$ at
$Q^2 \sim 1~{\rm GeV}^2 $, whereas
Ref.~\cite{Burkardt:2001dy} provides, still at very low scales $Q^2
\sim 0.4~{\rm GeV}^2 $, the value $ \langle x \rangle_q \approx
0.38 $.

For definiteness we adopt the values used in previous
works~\cite{Davidson:1994uv,Davidson:2001cc,RuizArriola:2002bp,
RuizArriola:2002wr}, namely that at $Q^2 = 4 {\rm GeV}^2 $ the valence
quarks carry $47 \%$ of the total momentum in the
pion~\cite{Sutton:1991ay,Capitani:2005jp}, {\em e.g.}, for $\pi^+$,
\begin{eqnarray} 
\langle x \rangle_v 
= 0.47(2) ,  \label{exprat}
\end{eqnarray} 
at $ Q^2 = 4~{\rm GeV}^2 $. At LO the scale turns out to be
\begin{eqnarray}
Q_0 = 313_{-10}^{+20} {\rm MeV},  
\label{Q0}
\end{eqnarray} 
where the value of $\Lambda_{\rm QCD}$ is provided in Eq.~(\ref{LambdaQCD}) and 
the error reflects the uncertainty in Eq.~(\ref{exprat}).
At such low scale the perturbative expansion parameter in the
evolution equations is large, ${\alpha(Q^2_0)}/({2\pi})=0.34$, which
makes the evolution very fast for the scales close to the initial
value. We return to this issue below.

\subsection{Evolution of PDF}

\begin{figure}[tb]
\includegraphics[width=0.4\textwidth]{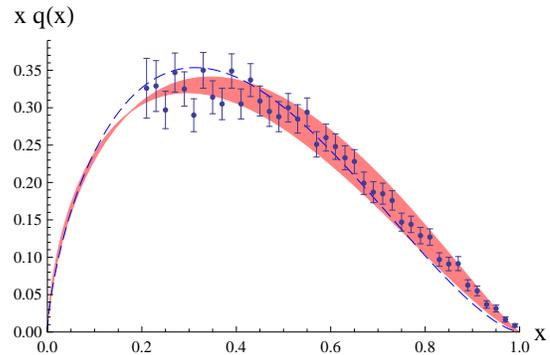}
\caption{(Color online) The quark model prediction for the valence
parton distribution (PDF) of the pion for a single quark (either $u$
or $\bar d$ for $\pi^+$) evolved to the scale of $Q=4$~GeV (band). The
width of the band indicates the uncertainty in the initial scale
$Q_0$, Eq.~(\ref{Q0}). The data points come from the analysis of the
Drell-Yan data from the E615 experiment \cite{Conway:1989fs}. The
dashed line shows the recent reanalysis of the original data made in
Ref.~\cite{Wijesooriya:2005ir}. \label{e615}}
\end{figure}

In Fig.~\ref{e615} we display the forward diagonal PDF of the pion, in
particular the quantity \mbox{$x q(x)=x H^{I=1}_q(x,0,0)$} obtained with
the LO QCD evolution up to $Q=4~{\rm GeV}$ from the quark-model initial
condition, $q(x,Q_0)=1$~\cite{Davidson:1994uv}. At leading order the
standard DGLAP evolution holds, which for the Mellin moments reads 
\begin{eqnarray}
\int_0^1 \, dx x^n q (x,Q) = \frac{1}{n+1}
\left(\frac{\alpha(Q)}{\alpha(Q_0) } \right)^{\gamma_n^{(0)}/(2
\beta_0)} \, ,
\end{eqnarray} 
where the anomalous dimensions for the vector vertex are given by 
\begin{eqnarray}
\!\!\!\!\!\!\!\!\! \gamma_n^{(0)} &=& -2 C_F \left[ 3 + \frac{2}{(n+1)(n+2)}- 4
\sum_{k=1}^{n+1} \frac1k \right],
\label{anom-dim}
\end{eqnarray}
with $C_F = (N_c^2 - 1)/(2N_c)$. The evolution equations can be solved
via the inverse Mellin transformation. In Fig.~\ref{e615} we confront
the result for $x q(x,Q)$ at the scale $Q=2$~GeV with the data at the
same scale from the E615 Drell-Yan experiment
\cite{Conway:1989fs}. The model results are represented with a band,
which reflects the uncertainty in the determination of the scale $Q_0$
in Eq.~(\ref{Q0}).  The quality of this comparison is impressive,
which shows that despite the rather embarrassingly low value of the
scale $Q_0$, the quark-model initial condition leads to fair
phenomenology.  The NLO evolution leads to small changes as compared
to the LO results~\cite{Davidson:2001cc}, in fact compatible with the
experimental uncertainties. The dashed line in represents the recent
reanalysis of the original E615 data made in
Ref.~\cite{Wijesooriya:2005ir}. We note that this result is also close
to the band generated by our quark-model calculations.

Moments of PDF's have been calculated on Euclidean
lattices~\cite{Best:1997qp}, yielding $\langle x \rangle_v = 0.3(1)$,
$\langle x^2 \rangle_v = 0.10(5)$ and $\langle x^3 \rangle_v =
0.05(1)$ at $Q=2.4 {\rm GeV}$. In Ref.~\cite{Capitani:2005jp} $\langle
x \rangle_v = 0.243(21)$ at $Q=2 {\rm GeV}$.

\begin{figure}[tb]
%%\vspace{-13mm}
%%\begin{center}
\includegraphics[angle=0,width=0.4\textwidth]{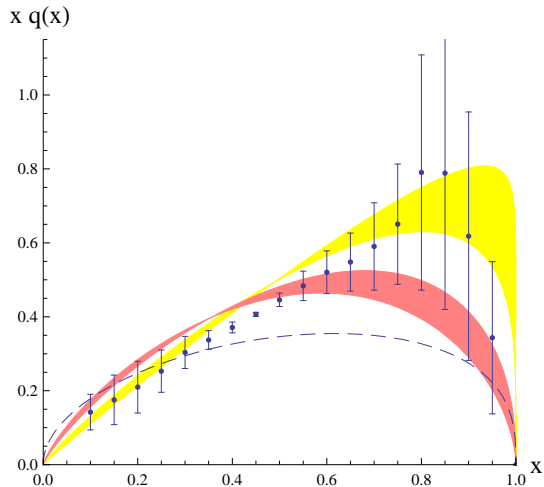}
%%\end{center}
\vspace{-1mm}
\caption{(Color online) The quark-model prediction for the valence
parton distribution function (PDF) of the pion for a single quark
(either $u$ or $\bar d$ for $\pi^+$) evolved to the scale $Q=0.5~{\rm
GeV}$ (darker band) and $Q=0.35~{\rm GeV}$ (lighter band). The width
of the bands indicates the uncertainty in the initial scale $Q_0$,
Eq.~(\ref{Q0}). The data come from the transverse lattice
calculations~\cite{Dalley:2002nj} and correspond to the scale $\sim
0.5$~GeV. The line shows the GRV \cite{Gluck:1998xa} parameterization
at $Q=0.5~{\rm GeV}$.}
\label{fig:pdf-transverse}
\end{figure}

The non-singlet PDF in the pion was also evaluated on the transverse
lattice~\cite{Dalley:2002nj} at the low renormalization scale $Q \sim
0.5~{\rm GeV}$.  In Fig~\ref{fig:pdf-transverse} we show our PDF
evolved to that scale (darker band) and to a lower scale of $0.35~{\rm
GeV}$ (lighter band).  We take the liberty of moving the scale, as its
determination on the lattice is not very precise.  As we see, the
agreement is qualitatively good if one considers the uncertainties of
the data, especially when the lower scale is used. We also show the
GRV98 parameterization \cite{Gluck:1998xa} (dashed line), which gives
somewhat lower PDF (except low values of $x$) compared to the lattice
data and our model results.

\subsection{Evolution of PDA}

\begin{figure}[tb]
%%\vspace{-13mm}
%%\begin{center}
\includegraphics[angle=0,width=0.4\textwidth]{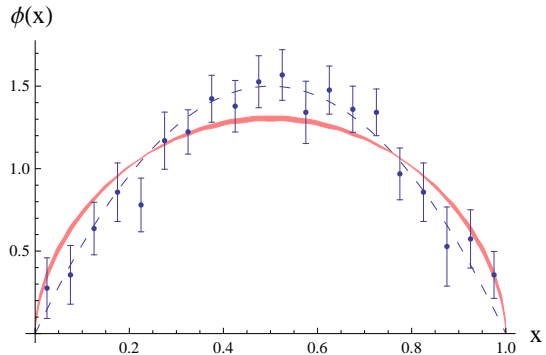}
%%\end{center}
\vspace{-1mm}
\caption{(Color online) The quark-model prediction for the pion distribution amplitude (PDA)
evolved to the scale $Q=2~{\rm GeV}$ (band) and compared to the E791 di-jet
measurement~\cite{Aitala:2000hb} after proper normalization of the data. The width of the band indicates the 
uncertainty in the initial scale $Q_0$, Eq.~(\ref{Q0}). We
also show the the asymptotic PDA, $\phi(x,\infty)=6x(1-x)$ (dashed line).}
\label{fig:pda-evol}
\end{figure}

PDA's have been intensely studied in the past in several contexts (see
e.g.~\cite{Anikin:1999cx,Dorokhov:2002iu,Bakulev:2003cs,Bakulev:2005cp} and
Ref.~\cite{Bakulev:2007jv} for a brief but comprehensive review). The
PDA of the pion \cite{RuizArriola:2002bp}, which can be related to the
isovector GPD through the soft pion theorem, Eq.~(\ref{eq:soft}), is
$\phi(x;Q_0) = 1 $, which holds at the quark model scale $Q_0$
(\ref{Q0})~\cite{RuizArriola:2002bp}. The evolved PDA can be expressed
in terms of the Gegenbauer
polynomials~\cite{Lepage:1980fj,Mueller:1994cn}
\begin{eqnarray}
\!\!\!\!\!\!\! \phi (x,Q) &=& 6x(1-x){\sum_{n=0}^\infty}'  C_n^{3/2} ( 2 x -1)
a_n (Q),
\label{eq:evolpda} 
\end{eqnarray}
where the prime indicates summation over even values of $n$ only.
Our inial condition yields~\cite{RuizArriola:2002bp}
\begin{eqnarray}
\!\!\!\!\!\!\!\! a_n (Q)&=& \frac23 \frac{2n+3}{(n+1)(n+2)} \left(
\frac{\alpha(Q)}{\alpha(Q_0) } \right)^{\gamma_n^{(0)} / (2
\beta_0)},
\label{OurGeg}
\end{eqnarray}
where the anomalous dimension for the axial-vector vertex is the same as for the 
vector vertex, $\gamma_n^{(0)}$, given in
Eq.~(\ref{anom-dim}). The evolved PDA is shown in
Fig.~\ref{fig:pda-evol}, where it is compared to the E791 di-jet
measurement~\cite{Aitala:2000hb}. The normalization of the di-jet data is used as a
fit parameter. Besides this normalization the result is {\it parameter
free}. As we see the agreement with the E791 data is rather reasonable
with a $\chi^2 /{\rm DOF}=1.45$. Nonetheless, the asymptotic wave
function generates a yet better $\chi^2 /{\rm DOF}=0.45$. Note that in
our scheme such a extreme limit would correspond to taking $Q_0 = \Lambda_{\rm QCD}$.

The second Gegenbauer moment at the scale $Q=2.4~{\rm GeV}$ is
$a_2=0.12 $ to be compared with the value $a_2 =0.12 (3)$ based on the
analysis of the CLEO data of Ref.~\cite{Schmedding:1999ap} where it
was assumed that $a_n=0$ for $n > 4 $.

Further, the leading-twist
contribution to the pion transition form factor is, at the LO in the
QCD evolution~\cite{Lepage:1980fj}, equal to
\begin{eqnarray}
\frac{Q^2 F_{\gamma^* \to \pi \gamma} (Q) }{2 f} \Big|_{\rm twist-2} = 
\int_0^1 dx \frac{\phi (x,Q )}{6x(1-x)}
\end{eqnarray}  
The experimental value obtained in CLEO~\cite{Gronberg:1997fj} for the
full form factor is $ Q^2 F_{\gamma^* , \pi \gamma} (Q) / (2 f) =
0.83 \pm 0.12 $ at $Q^2 = {\rm (2.4 GeV)}^2 $. Our value for the
integral, $1.25 \pm 0.10 $, overestimates the experimental result by
about two standard deviations, but one should bare in mind  that higher
twist contributions  as well as NLO perturbative
corrections have been ignored.

The second $\xi$-moment ($\xi = 2x-1$), defined as
\begin{eqnarray}
\langle \xi^2 \rangle_Q &=& \int_0^1 dx \, \phi (x, Q) (2x-1)^2,
\end{eqnarray} 
has been computed on Euclidean lattices yielding $\langle \xi^2
\rangle = 0.286 (49) $~\cite{DelDebbio:2002mq}, $\langle \xi^2 \rangle
= 0.269 (39) $~\cite{Braun:2006dg}, $\langle \xi^2 \rangle = 0.278
(26) $~\cite{Donnellan:2007xr} from recent Euclidean lattice
calculations at the scale $Q= 1/a \sim 2.6 \pm 0.1 {\rm GeV}$, where
$a$ is the lattice spacing. Note that the asymptotic PDA would yield
$\langle \xi^2 \rangle = 1/5=0.20$. We get from the quark model
$\langle \xi^2 \rangle= 0.244(4) $ for that scale.

\begin{figure}[tb]
%%\vspace{-13mm}
%%\begin{center}
\includegraphics[angle=0,width=0.4\textwidth]{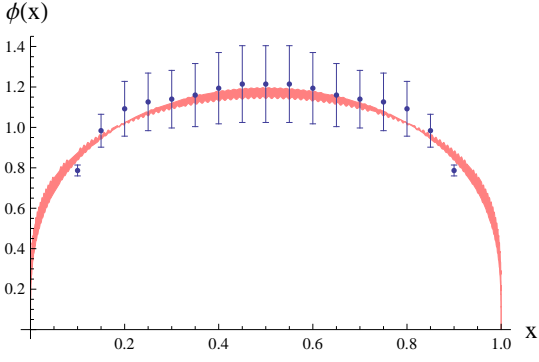}
%%\end{center}
\vspace{-1mm}
\caption{(Color online) The quark-model prediction for pion distribution amplitude (PDA) evolved to the scale 
$Q=0.5 {\rm GeV} $ (band) compared to the transverse lattice data~\cite{Dalley:2002nj}, corresponding to the scale 
$\sim 0.5$~GeV. The width of the band corresponds to the 
uncertainty in the initial scale $Q_0$, Eq.~(\ref{Q0}).}
\label{fig:pda-transverse}
\end{figure}

In Fig.~\ref{fig:pda-transverse} we compare our model prediction for
the PDA (band) to the transverse lattice data~\cite{Dalley:2002nj} at
the scale $Q=0.5$~GeV.  A good agreement is observed.

One of the most surprising aspects is that many of these results can
be obtained from the integral relation between PDF's and PDA's at a
given scale, established in Ref.~\cite{RuizArriola:2002bp}. The
relation allows to {\it predict} $\phi(x,Q)$ from $V(x,Q)$ as
parameterized, {\em e.g.}, by the Durham group~\cite{Sutton:1991ay}.
The method works for quark models, where $V(x,Q_0)=\phi(x,Q_0)=1$ at
some scale $Q_0$.

\subsection{End-point behavior\label{sec:end}}

\begin{figure}[tb]
\subfigure{\includegraphics[width=0.43\textwidth]{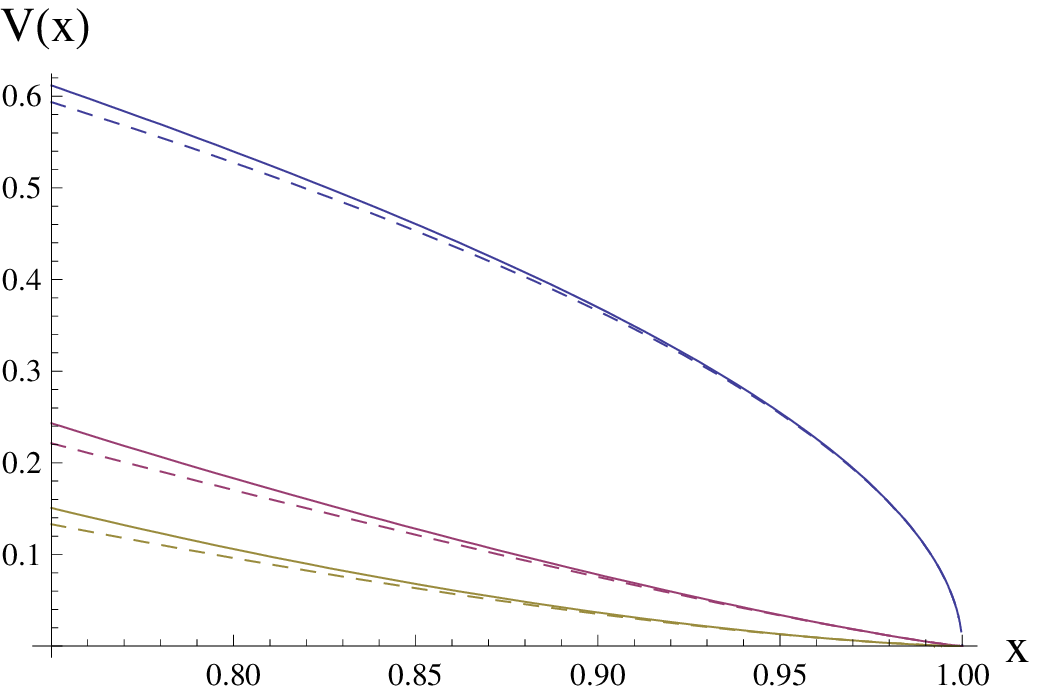}} \hspace{-10mm}
\subfigure{\includegraphics[width=0.43\textwidth]{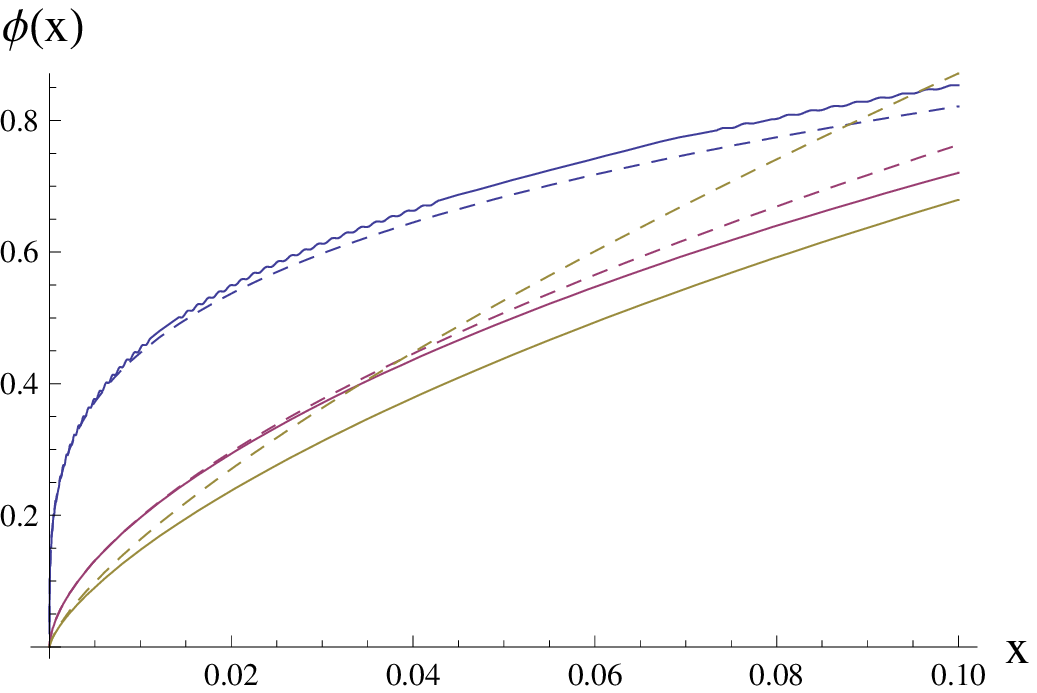}} \\ \vspace{-4mm}
\caption{(Color online) Comparison of the full result of the QCD
evolution (solid lines) to the asymptotic formulas near the end-points
(dashed lines). Top - valence PDF, bottom - PDA. The curves from top
to bottom are for $Q=0.5$, $2.4$, and $10$~GeV, respectively. The
initial scale $Q_0$ is taken from Eq.~(\ref{Q0}).
\label{fig:asym}}
\end{figure}

The results at the quark-model 
scale exhibit discontinuity at $x=0,1$, as $V(x,Q_0)=\phi(x,Q_0)=1$. 
An important feature of the evolution is that it cures the end-point behavior of the PDF's and PDA's
\cite{RuizArriola:2002bp,RuizArriola:2002wr,RuizArriola:2004ui}. 
Using the Mellin-moments formulation of the LO DGLAP evolution it can be shown 
that if $V(x,Q_0) \sim c (1-x)^p$ near $x=1$ 
then
\begin{eqnarray} 
V(x,Q) \sim c (1-x)^{p + 8 r(Q_0,Q)}, \qquad x\to 1, \label{endPDF}
\end{eqnarray}
where we have introduced the short-hand notation
\begin{eqnarray}
r(Q_0,Q)=\frac{C_F }{2\beta_0} \log \frac{\alpha(Q_0)}{\alpha(Q) }
\end{eqnarray} 
and $c$ and $p$ are some constants. Specifically, in our quark-model
case $c=1$ and $p=0$.  The prefactor can also be obtained, yielding
the more accurate formula \cite{RuizArriola:2004ui}
\begin{eqnarray} 
V(x,Q) \sim \frac{e^{2(3-4\gamma)r(Q_0,Q)}}{\Gamma(1+8
r(Q_0,Q))}(1-x)^{8 r(Q_0,Q)},\nonumber \\
\end{eqnarray}
where $x\to 1$, $\gamma$ denotes the Euler-Mascheroni constant, and
$\Gamma$ is the Euler gamma function.  At $Q > 1 {~\rm GeV}$ the
exponent of $1-x$ is a function weakly dependent on $Q$. The explicit
forms for several values of $Q$ and $Q_0$ from Eq.~(\ref{Q0}) are
\begin{eqnarray} 
V(x, 0.5 {~\rm GeV} ) &\sim& 1.23 (1-x)^{0.53}, \nonumber \\
V(x, 2.4 {~\rm GeV} ) &\sim& 1.13 (1-x)^{1.17}, \nonumber \\
V(x, 10 {~\rm GeV} ) &\sim& 1.00 (1-x)^{1.45} .
\end{eqnarray} 
These are compared to the full result in the top panel of Fig.~\ref{fig:asym}. 

Similarly, for the PDA evolved with the help of the Gegenbauer
polynomials (see Appendix \ref{app:end-point} for the derivation), one can show
that~\footnote{The corresponding formula in
Ref.~\cite{RuizArriola:2002bp} has a mistake.}
\begin{eqnarray} 
\phi(x, Q) &\sim& - \frac{\Gamma(-4 r(Q_0,Q))}{\Gamma(4r(Q_0,Q))}e^{2(3-4\gamma)r(Q_0,Q)} \times \nonumber \\
&&x^{4 r (Q_0,Q)}, \qquad x\to 0, \label{endPDA}
\end{eqnarray} 
and a symmetric expression when $x \to 1$ with $x$ replaced by $1-x$.
With several explicit values for $Q$ we have
\begin{eqnarray} 
\phi(x, 0.5 {~\rm GeV} ) &\sim& 1.51 \,x^{0.26}, \nonumber \\
\phi(x, 2.4 {~\rm GeV} ) &\sim& 2.95 \,x^{0.59}, \nonumber \\
\phi(x, 10 {~\rm GeV} ) &\sim& 4.65 \,x^{0.73} ,
\end{eqnarray} 
all for $Q_0$ from Eq.~(\ref{Q0}).  These asymptotic forms are
compared to the full result of the QCD evolution in the bottom panel
of Fig.~\ref{fig:asym}.  We note that the range of validity of the
approximation (\ref{endPDA}) shrinks closer and closer to the end point as $Q$ is
increased.  This must be so, as at $Q\to \infty$ the asymptotic form
$6x(1-x)$ sets in in the whole range of $x$, while the power
$r (Q_0,Q)$ increases indefinitely with $Q$.

It should be noted that the behavior of Eqs.~(\ref{endPDF},\ref{endPDA}), exhibiting the desired 
continuity of the functions at the end-points, is achieved already at values of $Q$ {\em infinitesimally} larger than $Q_0$.
Thus the QCD evolution heals the end-point problem immediately, at any $Q>Q_0$. Such a phenomenon is linked to the nonuniform 
convergence of the Mellin or Gegenbauer functional series near the end-points for the PDF and PDA, 
respectively.

\subsection{Evolution of diagonal GPD in the impact-parameter space}

The impact-parameter dependence quoted in Eq.~(\ref{eq:impact0}) at the
quark model point~\cite{Broniowski:2003rp} not only satisfies the
model independent relation~(\ref{impact-ff}) but after proper smearing
over plaquettes and DGLAP evolution qualitatively reproduces both the Bjorken $x$ and
impact parameter dependence when compared to transverse lattice
results~\cite{Dalley:2003sz,Dalley:2004rq} at the rather low scale
$Q \sim 0.5 {\rm GeV}$. This is a remarkable finding, since the transverse
lattice at such low scales should incorporate non-perturbative
evolution effects if they happened to be important.
Details can be found in Ref.~\cite{Broniowski:2003rp}.

\subsection{Discussion}

To sumarize this Section, the low value of the renormalization scale
$Q_0$ deduced from the LO perturbative evolution of the momentum
fraction, complies surprisingly well with a wealth of fragmentary
information for the non-singlet partonic distributions both on the
experimental side as well as compared to Euclidean lattices at $Q \sim
2 {\rm GeV}$ and transverse lattices at $Q \sim 0.5 {\rm GeV}$.  This
provides some confidence on applying a similar strategy to the
evolution of non-singlet GPD's as we do in the next section. Of
course, it would be of great help to have Euclidean lattices at small
renormalization scales, such that some of the non-perturbative
evolution could be explicitly seen. Unfortunately, the transition from
the intrinsically non-perturbative lattice regularization to the
perturbative $\overline{\rm MS}$ regularization scheme requires high
scales, so such a calculation seems hardly viable. Transverse lattices
do not suffer from this drawback, as these are non-perturbative
calculations at low scales~\cite{Burkardt:2001jg}. We are in
qualitative agreement also with these lattice calculations, which
probe the evolution in a region where it might potentially be highly
non-perturbative. Our analysis agrees within uncertainties with a
picture where the main non-perturbative feature of the valence quark
contribution is provided by the initial condition.  In any case, as
shown in the LO and NLO analysis of Ref.~\cite{Davidson:2001cc}, the
sea-quark and gluon PDF's from chiral quark models are less properly
reproduced. This might be improved if some non-singlet either sea or
gluonic model contributions could be provided at the model scale,
$Q_0$. Despite the efforts all over the years the problem of
determining the non-perturbative gluon content in a hadron at low
scales has remained unresolved.  These provisos should be taken in
mind when evolving the singlet GPD's in our scenario.

\begin{figure*}[tb]
\subfigure{\includegraphics[width=0.5\textwidth]{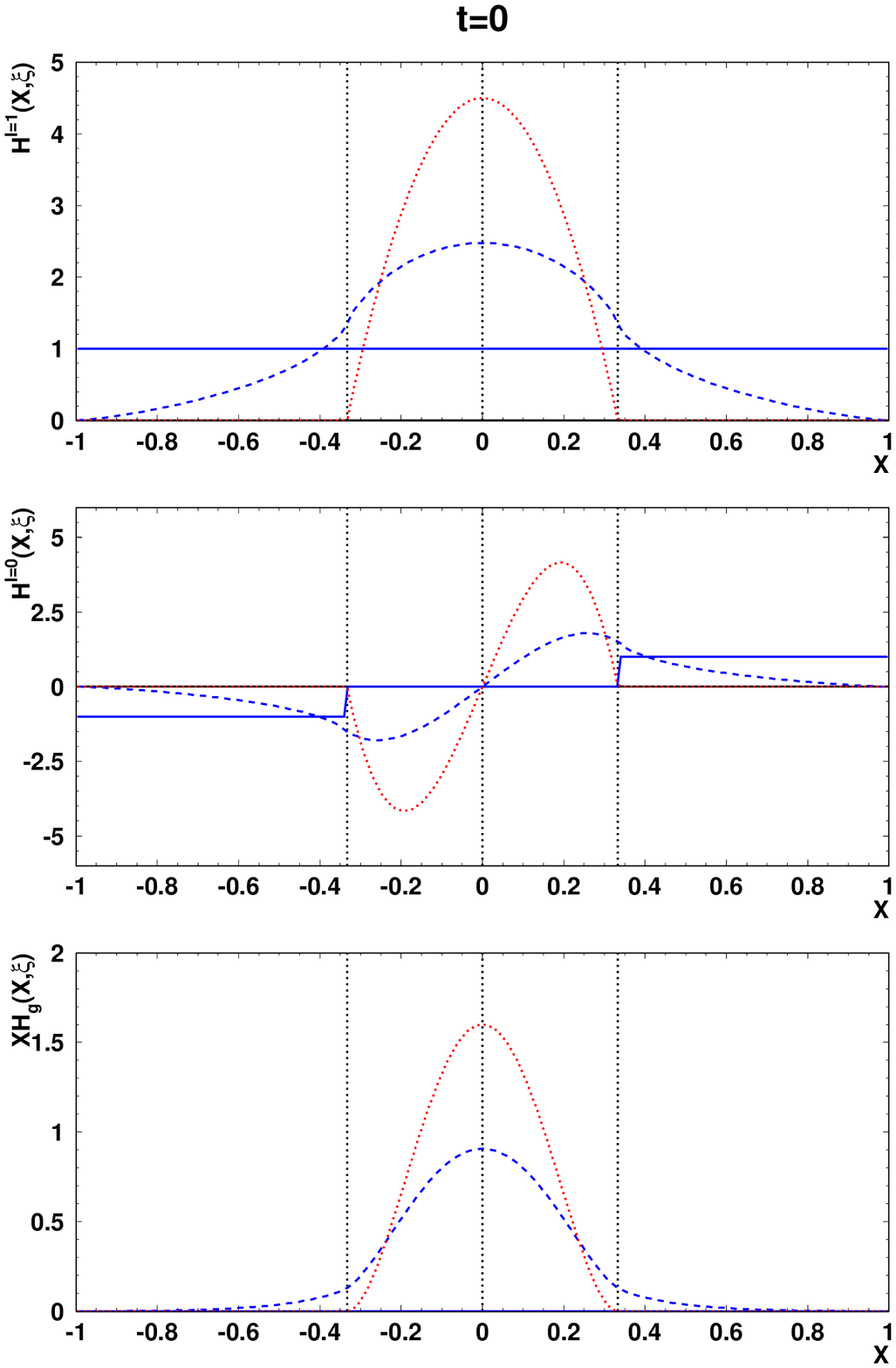}} \hspace{-10mm}
\subfigure{\includegraphics[width=0.5\textwidth]{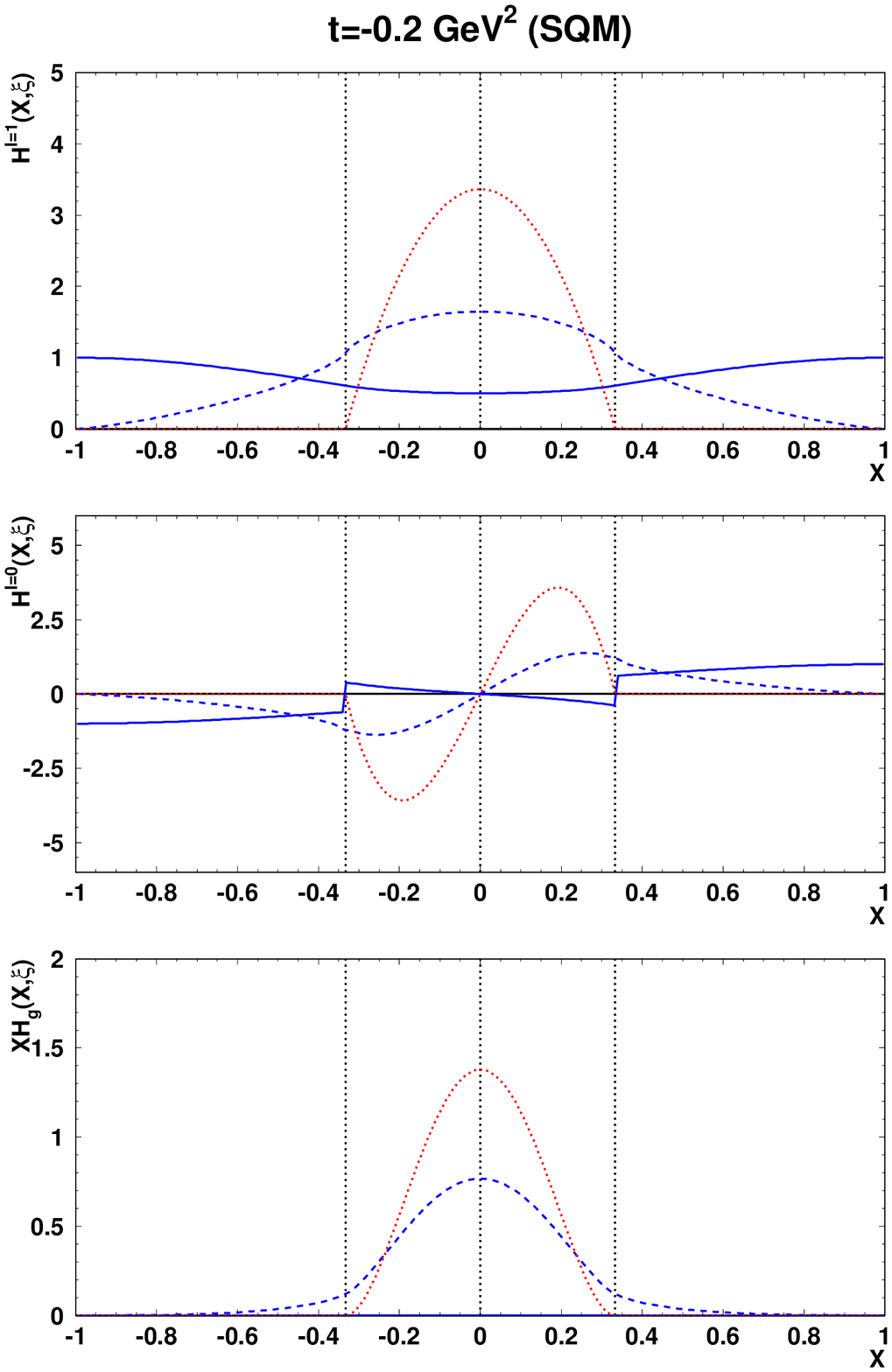}} \\ \vspace{-4mm}
\subfigure{\includegraphics[width=0.5\textwidth]{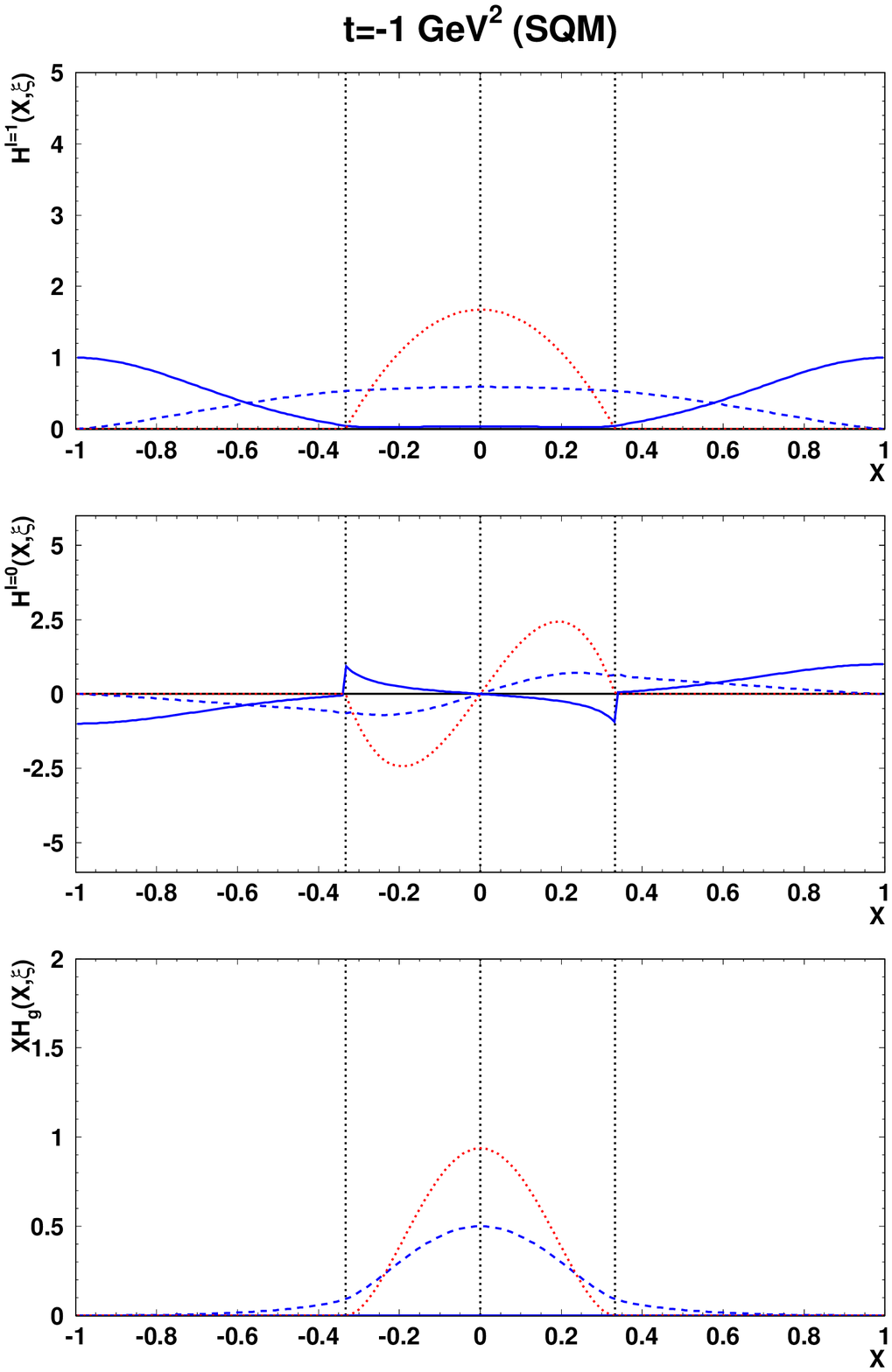}} \hspace{-10mm}
\subfigure{\includegraphics[width=0.5\textwidth]{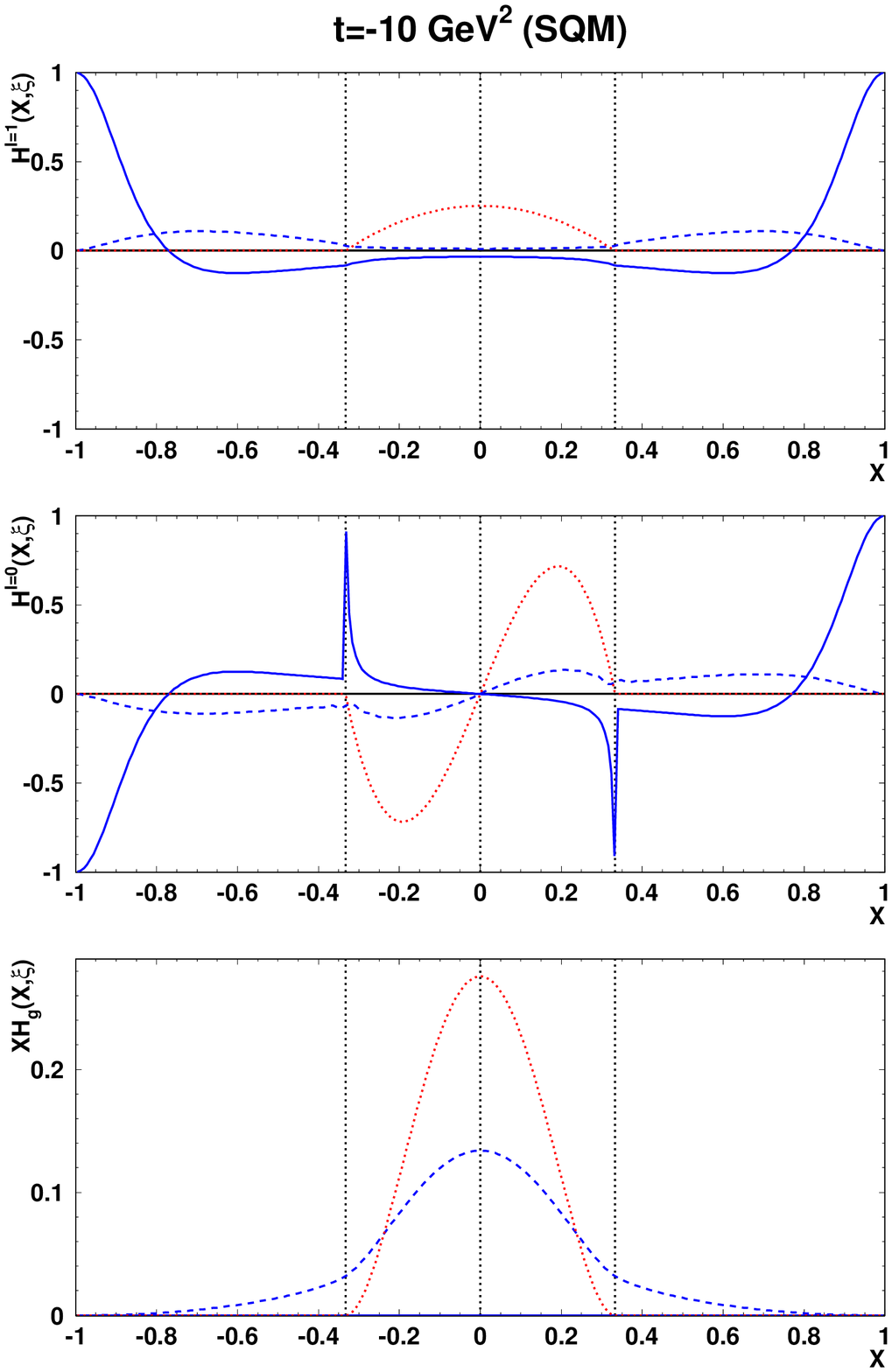}}
\caption{(Color online) Results of the LO QCD evolution
from the SQM initial condition for several values of $t$ and
$\xi=1/3$. Solid - initial condition at the quark-model scale, dashed
- evolution to $Q^2=(4 {\rm GeV})^2$, dotted - asymptotic form, $Q^2
\to \infty$.
\label{resevsqm}}
\end{figure*}

\section{The QCD evolution of GPD's \label{sec:evol}}

The explicit form of the LO QCD evolution equations for the GPD's can
be found in
\cite{Mueller:1998fv,Ji:1996nm,Radyushkin:1997ki,Blumlein:1997pi,GolecBiernat:1998ja,Kivel:1999wa,Kivel:1999sk}.
In this paper we solve them with the numerical method developed in
\cite{GolecBiernat:1998ja}, based on the Chebyshev polynomial
expansion.

As extensively discussed in the previous section, perturbative QCD
brings in the issue of evolution equations for the GPD's. Similarly to
the more familiar case of the PDF, the QCD interactions of massless
partons lead to collinear divergences which are factored out and
absorbed into the GPD's.  As a result, the GPD's become dependent on a
factorization (renormalization) scale $\mu$, usually identified with
the hard scale, $\mu=Q$. Thus, in general, the GPD's are functions of
four variables, $H=H(X,\xi,t,Q^2)$, with the kinematic constraints
$|X|\le 1$ and $0\le\xi\le 1$.  The renormalization group equations
which govern the dependence of GPD's on $Q^2$ are described in detail
{\em e.g.} in Ref.~\cite{GolecBiernat:1998ja}. The form of these
equations depends on the asymmetry parameter $\xi$, which defines two
regions: the Efremov-Radyushkin-Brodsky-Lepage (ERBL) region for
$|X|\le\xi$, and the Dokshitzer-Gribov-Lipatov-Altarelli-Parisi
(DGLAP) region for $|X|\ge \xi$.

An important feature of the GPD evolution, which makes it more
complicated than in the case of PDF or PDA, is that the evolution
equations in the ERBL region depend on the values of GPD's in the
DGLAP region. The converse is not true, the evolution in the DGLAP
region is not influenced by the the ERBL region.

Asymptotically, for $Q^2\to \infty$, the GPD's tend to the asymptotic forms
which are concentrated in the ERBL region only. In particular,
for $|X|<\xi$ we have 
\begin{eqnarray}
\label{eq:asymptotics}
&&H^{I=1}= \frac{3}{2\xi} \left(1-\frac{X^2}{\xi^2}\right) F_V(t) 
\\
&&H^{I=0}= (1 - \xi^2) \frac{15}{4 \xi^2} \frac{N_f}{4 C_F + N_f} \frac{X}{\xi}\left(1-\frac{X^2}{\xi^2}\right)\theta(t) \nonumber
\\
&&X H_g= (1 - \xi^2)  \frac{15}{16 \xi} \frac{4C_F}{4C_F + N_f} \left(1-\frac{X^2}{\xi^2}\right)^2 \theta(t)\nonumber
\end{eqnarray} 
while the GPD's vanish for $|X|\ge \xi$.
The proportionality constants reflect the normalization of the GPD's
at the initial scale $Q_0$, as the following charge- and momentum-conservation sum rules
are preserved by the evolution
\begin{eqnarray}
&&\int_{-1}^{1}dX\,H^{I=1}(X,\xi,t,Q^2)=2F_V(t),\\
&&\int_{-1}^{1}dX\,\left( X H^{I=0}(X,\xi,t,Q^2)+X H_g(X,\xi,t,Q^2)\right) \nonumber \\
&&=(1-\xi^2) F_S(t), \nonumber
\end{eqnarray}
in accordance to Eq.~(\ref{norm},\ref{norm2}).

\begin{figure}[tb]
\includegraphics[width=0.52\textwidth]{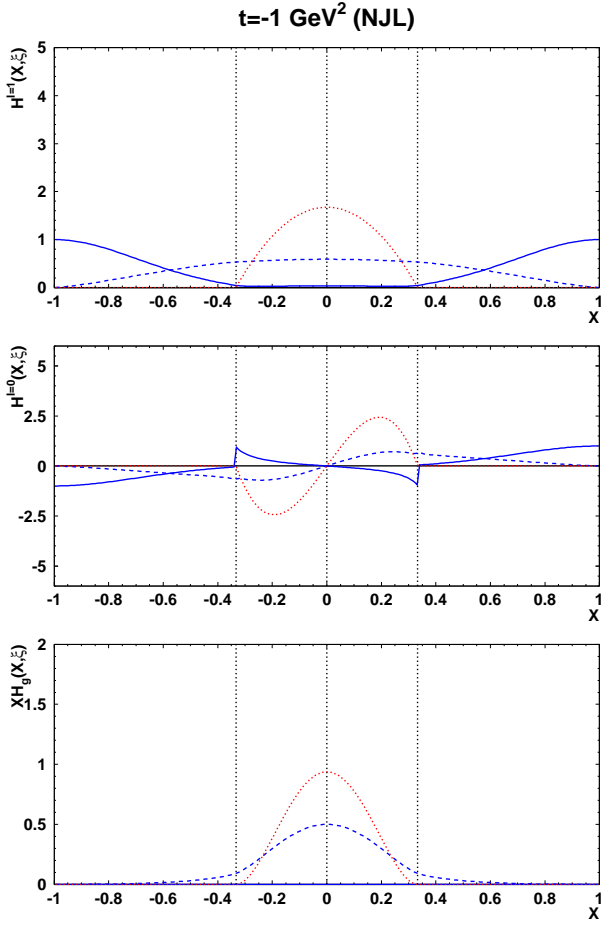}
\caption{(Color online) Same as bottom left panel of Fig.~\ref{resevsqm} for the NJL model. \label{resevnjl}}
\end{figure}

The results of the LO evolution from the SQM initial condition at the
scale $Q_0$ up to $Q=4~{\rm GeV}$ and $\xi=1/3$ are shown in
Figs.~\ref{resevsqm}. In each set of the four plots, corresponding to a
different value of $t$, we show the quark non-singlet $H^{I=1}$ (top) and singlet (middle)
$H^{I=0}$ distributions together with the gluon $H_g$ (bottom), conventionally multiplied by $X$.
We have chosen the sample value $\xi=1/3$, since the results are
qualitatively similar for other values of $\xi$.  The solid lines show
the initial condition at the quark-model scale $Q_0$ of
Eq.~(\ref{Q0}), the dashed lines show the result of the LO QCD
evolution to the scale $Q=4~{\rm GeV}$, and the dotted lines
show the asymptotic forms at $Q \to \infty$ given in
Eq.~(\ref{eq:asymptotics}). As the value of $-t$ is increased, the magnitudes
of the curves becomes lower, conforming to the sum rules
(\ref{norm},\ref{norm2}).  We note that the evolution smooths out the
original distributions, in particular, the discontinuities at the
end-points, $X=\pm 1$, and at the ERBL-DGLAP 
matching points $X=\pm \xi$ disappear for the isosinglet GPD.

The results for the NJL model are very similar to the case of SQM. In
Fig.~\ref{resevnjl} we show them for $t=-1~{\rm GeV}^2$ and $\xi=1/3$.
This similarity between the models is a sheer reflection of the
numerical similarity in the initial condition, {\em cf.}
Fig.~\ref{resevsqm} for $t=-1~{\rm GeV}^2$ and Fig.~\ref{resevnjl}.

\begin{figure}[tb]
\includegraphics[width=0.52\textwidth]{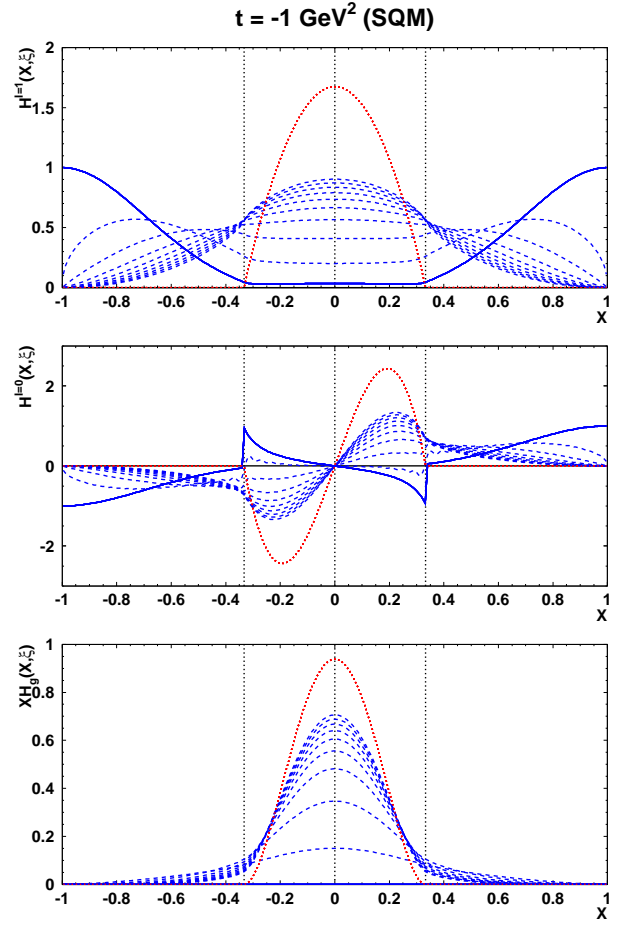}
\caption{(Color online) Same as Fig.~\ref{resevsqm} for subsequent evolution scales: $Q^2=0.1,1,10,10^2,\dots,10^8$~GeV${}^2$. Higher $Q^2$ 
gives higher magnitude of the curves in the ERBL region. \label{resQ}}
\end{figure}

In Fig.~\ref{resQ} it is shown how slow the evolution is in reaching
the asymptotic forms of the GPDF's.  The evolution is fastest at low
values of $Q$, where the coupling constant is large, and it
immediately pulls down the end-point values to zero. Then, the
strength gradually drifts from the DGLAP regions to the ERBL region.
Yet, the approach to the asymptotic form is very slow, with the tails
in the DGLAP region present. The highest $Q^2$ displayed in the figure
is $10^8$~GeV${}^2$ and the asymptotic form is reached at
``cosmologically'' large values of $Q$, which are never achieved
experimentally. Thus, the only way to approach the asymptotics would
be to start from initial conditions which are already close to it.  We
also observe that a larger $-t$ more strength of the quark GPDs
resides in the DGLAP region. This feature reflects the shape of the
initial condition, which inhibits the strength in the ERBL
region. Evolution up to $Q=2$~GeV retains this behavior, which
gradually disappears as $Q \to \infty$ where all the strength settles
in the ERBL region.

We note that the desired vanishing of the GPD's at the end-points $X=\pm 1$ is achieved due to the QCD evolution, 
similarly to the results presented in Sec.~\ref{sec:end}. Also, evolution leads to continuity at the DGLAP-ERBL boundary,
$X=\pm \xi$. These features are achieved at scales $Q$ infinitesimally above $Q_0$.

\section{Conclusions}
\label{concl}

We summarize our main points. In the present paper we have dealt with
the determination of the leading twist
GPD's of the pion in field theoretic chiral quark models. We have
done so with the help of an efficient method using the
$\alpha$-representation of the quark propagators and an extensive use of
double distributions. Our calculation incorporates the necessary
$D$-terms required by polynomiality and dimensional analysis. In the
chiral limit, we have been able to determine  explicit
analytic formulas for the pion GPD's.  All {\it a priori} properties which ought to be satisfied on
general principles, namely polynomiality,
positivity, proper support, soft-pion theorems, sum rules and
normalization are indeed fulfilled explicitly by our model
calculation. Although one might superficially think that these
properties should be trivially satisfied, the fact that one deals with
regularization or momentum dependence makes the fulfillment of those
properties less obvious, and in fact many calculations violate some
property.  A key ingredient in our approach has been a scrupulous
treatment of regularization in conjunction with electromagnetic and
chiral Ward-Takahashi identities. Our results for the pionic GPD's in the NJL 
model agree with Ref.~\cite{Theussl:2002xp}.

In the two chiral quark models considered, NJL and SQM, we have found results looking
alike since the models are mainly fixed by the pion charge form factor
which in both cases looks very similar. In addition, we have determined
the pion gravitational form factor entering the momentum-conservation sum rule. 
The outcoming GPD's are not
$t$-factorizable, an assumption which is being extensively used in
phenomenological approaches, not based on consistent dynamical
calculations.

However, with all those desirable properties fulfilled,
one must undertake the ERBL-DGLAP QCD evolution in order to relate the
model results to experimental high-energy data where higher-twist
contributions to the GPD's can be disregarded. Likewise, a comparison
to lattice results of the twist-2 GPD's requires specification of a
running scale. This aspect of the calculation is most frequently
ignored in dynamical model calculations, and particularly in chiral
quark models. A practical comparison to either experiment or lattice
can be be achieved by matching the momentum fraction of the QCD
evolved quark model to the experimental or lattice-extracted
result. In practice, the LO perturbative evolution is used with the result
that the low-energy quark-model scale is very low. Nevertheless, once this
is fixed the GPD's are uniquely determined. We have confronted our
predictions with all available information extracted either from
experiment or lattice both transverse or Euclidean. The experimental data include
the Fermilab E615 and E791 measurements of PDF and PDA of the pion,
respectively, and CLEO measuraments on the pion transition form factor.
The reasonable overall agreement to all these data corresponding to
quite different kinematical situations should be stressed. This fact provides
some confidence on our predictions of the non-singlet leading twist
GPD's. This also applies to the pion-photon Transition Distribution Amplitude, 
determined recently in quark models \cite{Broniowski:2007fs}. 

\begin{acknowledgments}
We warmly thank Sasha Bakulev and Simon Dalley for providing us with
the experimental and transverse lattice data, respectively.
\end{acknowledgments} 

\appendix

\section{The $\alpha$-representation evaluation of the two- and three-point functions \label{sec:alpha}}

In studies based on quark models the simplest way to obtain the GPD's
and the double distributions is through the use of the
$\alpha$-representation for the scalar propagators.  The advantage of
this representation over other popular Feynman parameterizations of
one-loop functions relies in the fact that the $\delta(k \cdot n -x)$
function constraining the loop integration is also naturally written
in terms of an integral of an exponential. This allows for maintaining
the explicit Lorentz covariance throughout the calculation. In our
scheme one does not have to start with the somewhat cumbersome moments
in $k \cdot n$, and then ``invert'' the result, as is frequently done.
The method used in this paper leads to well-defined and very simple
algebra and reproduces the double distributions in chiral quark models
from the literature. Also, in our approach the otherwise subtle effect
of the emergence of the $D$-terms follows in a clear way just from the
Feynman diagrams.

Below we derive basic integrals appearing later on in the evaluation
of the GPD's.  Calculations of this Appendix are made in the Euclidean
space.  We denote the Euclidean scalar propagators of particles of
mass $\omega$ as
\begin{eqnarray}
S_k=\frac{1}{D_k}=\frac{1}{k^2+\omega^2}=\int_0^\infty d \alpha e^{-\alpha (k^2+\omega^2)}, \label{eq:alpha}
\end{eqnarray}
where the RHS displays the $\alpha$-representation. 

\subsection{Two-point functions \label{sec:2p}}

Let us first consider the function $I(x,l \cdot n,0,l^2)$
corresponding to the definition (\ref{defIJ}) with the choice $l'=0$.
This two-point function with the constrained $k\cdot n$ integration
can be written as
\begin{widetext}
\begin{eqnarray}
I(x,\kappa,0,l^2)&=& \frac{4N_c w^2}{f^2} \int\frac{d^4 k}{(2\pi)^4} \delta(k \cdot n-x)S_k S_{k-l}=
\frac{4 N_c w^2}{f^2} \int\frac{d^4 k}{(2\pi)^4} \int \frac{d\lambda}{2\pi}
 e^{i \lambda (k \cdot n-x)}\!
\int_0^\infty \!\!\!\!\! da \int_0^\infty \!\!\!\!\!\!
db e^{-a (k^2+\omega^2)-b ((k-l)^2+\omega^2)} \nonumber \\
&=& \frac{4N_c w^2}{f^2} \int\frac{d^4 k'}{(2\pi)^4} \int_0^\infty \!\!\!\!\! da
\int_0^\infty \!\!\!\!\!db
\int \frac{d\lambda'}{2\pi} 
(a+b) e^{-(a+b)(k'^2+\omega^2)-\frac{ab}{a+b}l^2 + 
i \lambda'(b \kappa-(a+b)x)}\nonumber \\
&=& \frac{4N_c w^2}{f^2} \int\frac{d^4 k'}{(2\pi)^4} \int_0^\infty \!\!\!\!\! da
\int_0^\infty \!\!\!\!\!db \delta[b \kappa-(a+b)x]
(a+b) e^{-(a+b)(k'^2+\omega^2)-\frac{ab}{a+b}l^2},
\end{eqnarray}
\end{widetext}
where for brevity $\kappa=n\cdot l$, the shifted integration momentum is $k'=k+\frac{b}{a+b}l + \frac{i}{2} \lambda' n$, 
and $\lambda'=\lambda/(a+b)$. 
The $\delta[b \kappa-(a+b)x]$ function gives the constraint 
\begin{eqnarray}
x=\frac{b \kappa}{a+b}. \label{D1}
\end{eqnarray}
Since the integration variables are positive,  $a, b \ge 0$, it follows immediately that $x \in [0, \kappa]$ for $\kappa \ge 0$ and
\mbox{$x \in [-\kappa, 0)$} for $\kappa < 0$. This provides the proper 
support for $I(x,\kappa,0,l^2)$, which can be written generally as $\theta[x(\kappa-x)]$. 

One can decompose the $k'$ integration into two parts, 
\begin{eqnarray}
dk'^0 dk'^3=\pi dK^2, \;\;\; dk'^1 dk'^2=\pi du, \label{deco}
\end{eqnarray}
with $K^2=(k'^0)^2+(k'^3)^2$ and the ``transverse'' momentum $u=(k'^1)^2+(k'^2)^2$. 
Then  
\begin{widetext}
\begin{eqnarray}
I(x,\kappa,0,l^2)&=& \frac{N_c \omega^2 \theta[x(\kappa-x)]}{4\pi^2 f^2} \int_0^\infty \!\!\!\! du \int_0^\infty \!\!\!\! da 
\int_0^\infty \!\!\!\! db \, \delta[ b \kappa-(a+b)x] 
e^{-(a+b)(u+\omega^2)-\frac{a b}{a+b}l^2}  \label{D2} \\
&=& \frac{N_c \omega^2\theta[x(\kappa-x)]}{4\pi^2 f^2 |\kappa|} \int_0^\infty \!\!\!\!\! du \int_0^\infty 
db' e^{-b' \left [u+w^2 +\frac{x}{\kappa} \left ( 1-\frac{x}{\kappa} \right ) \right ]}
= \frac{N_c \omega^2 \theta[x(\kappa-x)]}{4\pi^2 f^2 |\kappa|}
\int_0^\infty \!\!\!\! du \frac{1}{u + w^2 + \frac{x}{\kappa} \left ( 1-\frac{x}{\kappa} \right )l^2},\nonumber
\end{eqnarray}
\end{widetext}
where $b'=b \kappa /x$. 
The integral over $u$ is logarithmically divergent, hence needs regularization, as expected.

The general function $I(x,\kappa,\kappa',(l-l')^2)$, where \mbox{$\kappa'=n \cdot l'$}, 
involves no extra work, as it can be obtained from the $l'=0$ case 
with the replacement
\begin{eqnarray}
k \to k-l', \;\; l \to l-l', \;\; x \to x - \kappa', \;\; \kappa \to \kappa - \kappa'.
\end{eqnarray}
This yields 
\begin{eqnarray}
&& I(x,\kappa,\kappa',(l-l')^2)
= \frac{N_c \omega^2 \theta[(x-\kappa')(\kappa-x)]}{4\pi^2 f^2 |\kappa-\kappa'|} \times \nonumber\\
&& \;\;\;\; \int_0^\infty \!\!\!\! du \frac{1}{u + w^2 + \frac{x-\kappa'}{\kappa-\kappa'} 
\left ( 1-\frac{x-\kappa'}{\kappa-\kappa'} \right )(l-l')^2}. \label{gen2}
\end{eqnarray}

An important consequence of Lorentz invariance is {\em polynomiality}
\cite{Ji:1998pc}.  We verify it by introducing the variable
$\nu=(x-\kappa')/(\kappa-\kappa')$, when
$I(x,\kappa,\kappa',(l-l')^2)$ becomes a function of $\nu$ devided by
$|\kappa-\kappa'|$. We obtain (assuming for definiteness $\kappa >
\kappa'$)
\begin{eqnarray} 
&& \int_{-1}^1 dx I(x, \kappa, \kappa',(l-l')^2) x^n = \nonumber \\
&& \;\;\;\;\; \int_{0}^1 d\nu f(\nu) [\kappa'+\nu (\kappa - \kappa')]^n, 
\end{eqnarray}
which results in a polynomial in $\kappa$ and $\kappa'$ of the order at most $n$.
The first few moments have the explicit form
\begin{widetext}
\begin{eqnarray}
&&\int_{-1}^1 dx I(x, \kappa, \kappa',\tau) =  \frac{N_c \omega^2 \tau}{4\pi^2 f^2} \int_0^\infty du
\frac{2 \log \left(\frac{\sqrt{4 A+1}-1}{\sqrt{4 A+1}+1}\right)}{\sqrt{4 A+1}}, \nonumber \\
&&\int_{-1}^1 dx I(x, \kappa, \kappa',\tau)\,x = \frac{N_c \omega^2 \tau}{4\pi^2 f^2} \int_0^\infty du
\frac{(\kappa+\kappa') \log \left(\frac{\sqrt{4 A+1}-1}{\sqrt{4 A+1}+1}\right)}{\sqrt{4 A+1}}, \nonumber \\
&& \int_{-1}^1 dx I(x, \kappa, \kappa',\tau)\,x^2 = \frac{N_c \omega^2 \tau}{4\pi^2 f^2} \int_0^\infty du \times \nonumber \\
&& \;\;\; \frac{(4 A+1) \left(\kappa -\kappa '\right)^2-\sqrt{4 A+1} \log \left(\frac{\sqrt{4 A+1}+1}{\sqrt{4 A+1}-1}\right) \left((2 A+1) \kappa ^2-4 A \kappa ' \kappa +(2 A+1)
   \left(\kappa '\right)^2\right)}{4 A+1},
\end{eqnarray}
\end{widetext}
where $A=(u+\omega^2)/\tau$ and $\tau=(l-l')^2$.

In the literature the $D$-term is by definition the two-point function
in the $t$-channel \cite{Polyakov:1999gs}.  It originates from the
diagram with the contact pion-quark term as well as from the reduced
three-point diagram, where by ``reduction'' one means the replacement
of $k^2$ and $k \cdot l$ pieces appearing in the numerator from the
trace factor, in terms of the inverse scalar propagators.  The
two-point functions in the $s$-channel (resulting from the reduction
of the three-point function) are traditionally treated as singular
parts of the double distributions.

\subsection{Three-point functions \label{sec:3p}}

For the three-point functions we proceed analogously, now with three
scalar propagators.  We need to take into account the kinematics of
the {\em direct} and {\em crossed} diagrams of Fig.~\ref{fig:diag}.
We first analyze in detail the the three-point function resulting from
the direct diagram (a), since the case of the crossed diagram is
obtained via a simple kinematic transformation. We have
\begin{widetext}
\begin{eqnarray}
J(x,q\cdot n,p \cdot n,q^2,p^2,p\cdot q)&=& \frac{4N_c \omega^2}{f^2} \! \int\frac{d^4 k}{(2\pi)^4} 
\delta(k \cdot n-x) S_k S_{k+q} S_{bk-p} \\ &=&
{4 N_c \omega^2}{f^2}\! \int\frac{d^4 k}{(2\pi)^4} \int \frac{d\lambda}{2\pi} e^{i \lambda (k \cdot n-x)} 
\int_0^\infty \!\!\!\!\! da \int_0^\infty \!\!\!\!\!
db \int_0^\infty \!\!\!\!\!
dc e^{-a (k^2+\omega^2) -b ((k+q)^2+\omega^2) - c ((k-p)^2+\omega^2) }. \nonumber
\end{eqnarray}
%\end{widetext}
Shifting the integration variable, $k'=k+(\beta q-\gamma p-i \lambda
n/2)/(\alpha+\beta+\gamma)$, and carrying over the $d^4k'$ integration
yields
%\begin{widetext}
\begin{eqnarray}
\!\!\!\!J&=& 
\frac{N_c \omega^2}{4\pi^2 f^2} \int_0^\infty \!\!\!\!\! da 
\int_0^\infty \!\!\!\!\!db 
\int_0^\infty \!\!\!\!\!dc \, \frac{1}{(a+b+c)^2}
\delta \left ( x - \frac{c p\cdot n - b q \cdot n}
{a+b+c} \right ) e^{-(a+b+c)\omega^2-
\frac{b (a+c)}{a+b+c}q^2 - 
\frac{c (a+b)}{a+b+c}p^2 -
\frac{2 b c}{a+b+c}p \cdot q}.
\end{eqnarray}
\end{widetext}
Next, we change the variables into
\begin{eqnarray}
s=a+b+c, \;\; y=\frac{b}{s}, \;\; z=\frac{c}{s}. 
\end{eqnarray}
Note that since $a, b, c \ge 0$, we get 
$0 \le y, z \le 1$ and also $y + z \le 1$.
The substitution and integration over $s$ yields
%\begin{widetext}
\begin{eqnarray}
J &=& \frac{N_c \omega^2}{4\pi^2 f^2} \times \label{bub} \\ && \int_0^1 \!\! dy 
\int_0^1 \!\! dz \frac{\theta(1-y-z)
\delta \left ( x - z p\cdot n + y q \cdot n \right )}
{\omega^2+ y(1-y)q^2 + z(1-z)p^2 + 2 y z p \cdot q}. \nonumber
\end{eqnarray}
We note that polynomiality is obvious from this form, as
multiplication by the power $x^n$ is equivalent to the multiplication
by the factor $(z p\cdot n - y q \cdot n)^n$.  In the chiral limit of
$m_\pi=0$ the first few moments are relatively simple:
\begin{widetext}
\begin{eqnarray}
&&\int_0^1 \!\!\!\! dx \, J = 
\frac{N_c \omega^2}{4\pi^2 f^2 t}  2 \left [ \arctan\left(\frac{\sqrt{t}}{\sqrt{4 w^2-t}}\right) \right]^2 , \\
&&\int_0^1 \!\!\!\! dx \, J\, x = \frac{N_c \omega^2}{4\pi^2 f^2 t^{3/2}}
2 \left(\sqrt{t} \left[  \arctan\left(\frac{\sqrt{t}}{\sqrt{4 w^2-t}}\right) \right]^2
-\sqrt{4 w^2-t} (\zeta -2) \arctan\left(\frac{\sqrt{t}}{\sqrt{4
   w^2-t}}\right)+\sqrt{t} (\zeta -2)\right), \nonumber \\
&&\int_0^1 \!\!\!\! dx \, J\, x^2 = \frac{N_c \omega^2}{4\pi^2 f^2 t^2}  \left [ 
t (\zeta  (\zeta +3)-7)+ \right . \nonumber \\ 
&& \;\;\;\;\; \left . \arctan \left(\frac{\sqrt{t}}{\sqrt{4 w^2-t}}\right) \left(\sqrt{t} \sqrt{4 w^2-t} (6-\zeta  (\zeta +2))+2
   \left(t-2 w^2 (\zeta -1)\right) \arctan\left(\frac{\sqrt{t}}{\sqrt{4 w^2-t}}\right)\right) \right ]. \nonumber
\end{eqnarray}
\end{widetext}

We can rewrite Eq.~(\ref{bub}) as
\begin{eqnarray}
{\cal F}(z,y)&=&\frac{N_c \omega^2}{4\pi^2 f^2} \frac{\theta(1-y-z)}
{\omega^2 - y(1-y-z) t - z(1-z)m_\pi^2}, \nonumber \\
J(x) &=& \int_0^1 \!\! d y \int_0^1 \!\! d z 
\delta \left ( x - z  - y \zeta \right ){\cal F}(z,y), \nonumber \\
\label{bubzy}
\end{eqnarray}
where we have used the kinematics (\ref{kin}). The curly ${\cal F}$ denotes the {\em double distribution}.

Let us denote
\begin{eqnarray}
\!\!\!\!\!\!\!\! {\cal D}={\omega^2 - y(1-y-z) t - z(1-z)m_\pi^2}. \label{curlyD}
\end{eqnarray}
For the GPD of the pion, due to the crossing symmetry, one may assume
$0 \le \zeta \le 1$. Next, we perform the $z$ integration, which sets
\begin{eqnarray}
z=x-y \zeta.
\end{eqnarray}
The distributions in Eq.~(\ref{bub})
give the following limits for the $y$ integration:
\begin{eqnarray}
&& \!\!\!\!\!\!\! J = \frac{N_c \omega^2}{4\pi^2 f^2} \left (
 \theta[x(\zeta-x)]\!\!\int_0^{\frac{x}{\zeta}} \!\!
+ \theta[(x-\zeta)(1-x)] \!\!\int_0^{\frac{1-x}{1-\zeta}} 
\right ) \frac{d y}{\cal D} ,\nonumber \\ && 
\label{Fp}
\end{eqnarray}
with the first term having the support $x \in [0,\zeta]$, and the
second $x \in [\zeta,1]$.  The function $F(x)$ is continuous, but the
derivative $dF(x)/dx$ is discontinuous at the points $x=0,\zeta, 1$.
The double distribution is {\em
M\"unchen-symmetric}~\cite{Mankiewicz:1997uy}, {\em i.e.}  ${\cal
F}(z,y)={\cal F}(z,1-y-z)$. This feature is related to the crossing
symmetry, holding for identical particles.

The result for the crossed diagram (see Fig.~\ref{fig:diag} is
obtained from the above result for the direct diagram with the
replacement $p \to -p - q$. Replacing correspondingly $x \to \zeta-x$
and performing the M\"unchen transformation~\cite{Mankiewicz:1997uy}
\begin{eqnarray}
z\to z, \;\;\; y \to 1-y-z, \label{mue}
\end{eqnarray}
we find that 
${\cal D}$ is {\em invariant} under this joint transformation. 
The function $\delta(x-z-y\zeta)$ is also invariant under these combined
two trasformations. Finally, the support
is invariant, since
\begin{eqnarray}
&&\theta(1-y-z) \theta[y(1-y)] \theta[z(1-z)] \to \nonumber \\
&& \;\;\;\;  \theta(y) \theta[(1-y-z)(y+z)] \theta[z(1-z)] \nonumber \\ 
&& \;\;\;\; 
= \theta(1-y-z) \theta[y(1-y] \theta[z(1-z)],    
\end{eqnarray}
where the equality in the above formula is an algebraic identity. 
Therefore the crossed diagram is related to the direct diagram as follows:
\begin{eqnarray}
J^{\rm crossed}(x,\zeta)=J^{\rm direct}(\zeta-x,\zeta).
\end{eqnarray}
The support of the crossed diagram reflects the support of the direct diagram, {\em i.e.}, $x \in [-1+\zeta,\zeta]$.

\section{The two- and three-point functions in the Spectral Quark Model}

According to the general rule, in SQM one append the formulas with the spectral integration
$\int_C d\omega \omega^2 \rho(\omega)$. The results below are for the meson-dominance model.

\subsection{The two-point function}

We assume $\kappa' \le \kappa$. The spectral integration yields 
\begin{eqnarray} 
&& I_{\rm SQM}(x,\kappa,\kappa',l^2)=\int_C d\omega \omega^2 \rho(\omega) I(x,\kappa,\kappa',l^2)  \nonumber \\
&& = \frac{\theta[(x-\kappa')(\kappa-x)]}{(\kappa-\kappa') \left [ 1+ 
4 \frac{x-\kappa'}{\kappa-\kappa'} \left ( 1-\frac{x-\kappa'}{\kappa-\kappa'} \right )\frac{l^2}{M_V^2} \right ]^{3/2}}, \label{s1}
\end{eqnarray}
where we have used the relation 
\begin{eqnarray}
M_V^2=24 \pi^2 f^2/N_c.
\end{eqnarray} 
The integration over $x$ yields the form factor
\begin{eqnarray}
\int dx  I_{\rm SQM}(x,\kappa,\kappa',l^2)= \frac{M_V^2}{M_V^2+l^2}.
\end{eqnarray}
In agreement with polynomiality, this form factor is independent 
of the value of $\kappa$ or $\kappa'$. Note that the vector-meson dominance is readily obtained.
Similarly, 
\begin{eqnarray}
\int dx  I_{\rm SQM}(x,\kappa,\kappa',l^2)\,x= \frac{M_V^2}{M_V^2+l^2}\frac{\kappa+\kappa'}{2}.
\end{eqnarray}

\subsection{The three-point function}

For simplicity in this Appendix we work in the chiral limit. In this case the double distribution becomes
\begin{eqnarray}
{\cal F}_\omega(z,y;t)&=& \frac{N_c \omega^2}{4\pi^2 f^2} \frac{\theta(1-y-z)}
{\omega^2 - y(1-y-z)t}, 
\end{eqnarray}
and the subsequent spectral integration yields (again we only multiply by $\omega^2$ and leave out 
other factors)
\begin{eqnarray}
{\cal F}_{\rm SQM}(z,y;t)&=&\int_C d\omega \rho(\omega) {\cal F}_\omega(z,y;t) 
\nonumber \\ &=& \frac{N_c}{4\pi^2 f^2} \frac{\theta(1-y-z)}
{\left ( 1 - \frac{4 y(1-y-z)t}{M_V^2} \right )^{5/2}}. 
\end{eqnarray}
Let us introduce the short-hand notation
\begin{widetext}
\begin{eqnarray}
\chi_2&=&\frac{2 (x-1) \left[ 3 (\zeta-1) M_V^2+t (x-1)^2\right ]}{\left[ (\zeta-1)
   M_V^2+t (x-1)^2\right]^2}, \nonumber \\
\chi_1&=& 
\frac{(x (\zeta -2)+\zeta ) \left(3 M_V^2 (\zeta -1) \zeta ^2+t \left(\left(\zeta ^2+8 \zeta -8\right) x^2+2 (4-5 \zeta ) \zeta  x+\zeta
   ^2\right)\right)}{\left((\zeta -1) M_V^2+t (x-1)^2\right)^2 \left({\zeta^2+\frac{4 t x (x-\zeta )}{M_V^2}}\right)^{3/2}} +\frac{1}{2}\chi_2.
\end{eqnarray}
\end{widetext}
Then, for the case $\zeta \ge 0$,
\begin{eqnarray}
J_{\rm SQM}(x,\zeta;t) &=& \left ( \theta[x(\zeta-x)] \chi_1 
% \right . \nonumber \\ &+& \left . 
+ \theta[(1-x)(x-\zeta)] \chi_2 \right ).\nonumber \\ \label{chi}
\end{eqnarray} 
The function satisties $J_{\rm SQM}(0,\zeta;t)=J_{\rm SQM}(1,\zeta;t)=0$. 
The value at the matching point $x=\zeta$ is
\begin{eqnarray}
J_{\rm SQM}(\zeta,\zeta;t)=
\frac{2 \left(3 M_V^2+t (\zeta-1)\right)}{\left(M_V^2+t (\zeta-1)\right)^2}. \label{ma}
\end{eqnarray}
The integration over $x$ produces a $\zeta$-independent (as required by polynomiality) 
form factor,
\begin{eqnarray}
\int_0^1 \!\! dx J_{\rm SQM}(x,\zeta;t)=
   \frac{2}{M_V^2-t}-\frac{\log 
   \left(1-\frac{t}{M_V^2}\right)}{t} . \label{ffF}
\end{eqnarray}
Similarly,
\begin{eqnarray}
\int_0^1 \!\! dx J_{\rm SQM}(x,\zeta;t)\, x=
   \frac{\zeta}{M_V^2-t}-\frac{\log 
   \left(1-\frac{t}{M_V^2}\right)}{t} . \label{ffF2}
\end{eqnarray}

For the special case of $t=0$ Eq.~(\ref{chi}) reduces to the very simple expression
\begin{eqnarray}
J_{\rm SQM}(x,\zeta;0) &=& \frac{6}{M_V^2} \left ( \theta[x(\zeta-x)] \frac{x}{\zeta}
\right . \nonumber \\ &+& \left . \theta[(1-x)(x-\zeta)] \frac{x-1}{\zeta-1} \right ),\nonumber \\ 
&& \label{chi2}
\end{eqnarray} 
which is a triangle of area $3/M_V^2=N_c/(8\pi^2 f^2)$.

For the case $\zeta=0$ we have 
\begin{eqnarray}
J_{\rm SQM}(x,0;t) &=& \frac{\left(3 M_V^2-t (1-x)^2\right) (1-x)}{\left(M_V^2-t (1-x)^2\right)^2}.
\end{eqnarray}

\section{The two- and three-point functions in the NJL model \label{app:njl}}

Operationally, the calculation in the NJL model with the regularization (\ref{prescr}) amounts to 
taking the generic expressions (\ref{gen2}) and (\ref{Fp}), replacing $\omega^2 \to M^2+\Lambda^2$ {\em in the denominators}, 
carry out the integrations, and finally applying (\ref{prescr}). We work in the chiral limit.

\subsection{The two-point function}

Through the use of Eq.~(\ref{f2njl}) we arrive immediately at the formula (for $\kappa \ge \kappa'$)
\begin{eqnarray}
I_{\rm NJL}(x,\kappa,\kappa',0)= \frac{\theta[(x-\kappa')(\kappa-x)]}{\kappa-\kappa'}. \label{Injl}
\end{eqnarray}

\subsection{The three-point function}

We find
\begin{widetext}
\begin{eqnarray}
&& J_{\rm NJL}(x,\zeta;m_\pi=0;\Lambda)=- \frac{N_c M^2}{8 \pi^2 f^2} \left [ \frac{2 \log \left(\frac{\sqrt{-t} (x-1)+\sqrt{-t (x-1)^2-4 (\zeta -1) \left(M^2+\Lambda ^2\right)}}{\sqrt{-t (x-1)^2-4 (\zeta -1)
   \left(M^2+\lambda ^2\right)}-\sqrt{-t} (x-1)}\right) \theta ((1-x) (x-\zeta ))}{\sqrt{-t} \sqrt{-t (x-1)^2-4 (\zeta -1)
   \left(M^2+\lambda ^2\right)}}+\right . \\&& \left . \frac{\left(\log \left(\frac{\sqrt{-t} (x-1)+\sqrt{-t (x-1)^2-4 (\zeta -1) \left(M^2+\lambda
   ^2\right)}}{\sqrt{-t (x-1)^2-4 (\zeta -1) \left(M^2+\lambda ^2\right)}-\sqrt{-t} (x-1)}\right)+\log \left(\frac{\sqrt{-t (x-1)^2-4
   (\zeta -1) \left(M^2+\lambda ^2\right)} \zeta +\sqrt{-t} (x (\zeta -2)+\zeta )}{\zeta  \sqrt{-t (x-1)^2-4 (\zeta -1)
   \left(M^2+\lambda ^2\right)}-\sqrt{-t} (x (\zeta -2)+\zeta )}\right)\right) \theta (x (\zeta -x))}{\sqrt{-t} \sqrt{-t (x-1)^2-4
   (\zeta -1) \left(M^2+\lambda ^2\right)}} \right ] , \nonumber
\end{eqnarray}
\end{widetext}
and 
\begin{eqnarray}
J_{\rm NJL}(x,\zeta;m_\pi=0)=\left . J_{\rm NJL}(x,\zeta;m_\pi=0;\Lambda)\right |_{\rm reg}.
\end{eqnarray}
Polynomiality follows from the fact that the expressions are derived from the double distributions (\ref{bubzy}) and the distributive
nature of the regularization prescription (\ref{prescr}). By {\em distributive} we mean that it is a sum over quark masses or the 
integral over $\omega$ of the formal expressions for $I$ and $J$.

\section{The gravitational form factors \label{app:grav}}

The gravitational form factors of the pion \cite{Donoghue:1991qv} are defined through the matrix element of the energy-momentum tensor,   
\begin{eqnarray}
&& \!\!\!\!\!\! \langle \pi^b(p+q) \mid \theta^{\mu \nu}(0) \mid \pi^a(p) \rangle = \\ && 
\!\!\!\!\!\!\! \frac{\delta^{ab}}{2}\left (g^{\mu \nu}q^2- (q^\mu q^\nu)
\theta_1(q^2)+ (2p+q)^\mu (2p+q)^\nu \theta_2(q^2) \right ). \nonumber
\end{eqnarray}
They satisfy the low energy theorem $\theta_1(0) - \theta_2(0) = {\cal
O} (m_\pi^2) $~\cite{Donoghue:1991qv}. The leading-$N_c$ quark-model
evaluation amounts to computing the diagrams of Figs.~\ref{fig:diag}
and \ref{fig:diag2} with the pion gravitational vertex
\begin{eqnarray}
&&\theta^{\mu \nu}(k+q,k)=\frac{1}{4} \left ( (2k+q)^\mu \gamma^\nu + (2k+q)^\nu \gamma^\mu \right ) \nonumber \\
&& - \frac{1}{2} g^{\mu \nu} \left ( 2 \slashchar{k} + \slashchar{q} - \omega \right ).  
\end{eqnarray}
The results of the calculation in SQM is Eq.~(\ref{ffSQMg}). Equation
(\ref{norm2}) follows from considering the matrix element of $n_\mu
\theta^{\mu \nu} n_\nu$. Then
\begin{eqnarray}
&& \langle \pi^b(p+q) \mid n_\mu \theta^{\mu \nu}(0) n_\nu \mid \pi^a(p) \rangle = \\ && 
\delta^{ab} \frac{1}{2}\left [
\zeta^2 \theta_1(q^2)+ (2-\zeta)^2 \theta_2(q^2) \right ]. \nonumber
\end{eqnarray}
The vertex becomes 
\begin{eqnarray}
&&n_\mu \theta^{\mu \nu}(k+q,k)n_\nu =(x-\zeta/2) \gamma \cdot n.  
\end{eqnarray}
We notice it is the same vertex as in the evaluation of the GPD's multiplied by $(x-\zeta/2)$.
Upon passing to the symmetric notation Eq.~\ref{norm2} follows.

\section{End-point analysis for the PDA}
\label{app:end-point}

Here we derive the formulas used in the main text of
Sect.~\ref{sec:end} for the PDA and correct a mistake 
in expressions of our previous
work~\cite{RuizArriola:2002bp,RuizArriola:2002wr}.  Right at the end-points,
$ x= 0,1$, the series (\ref{eq:evolpda}) diverges since $ C_{2k}^{3/2}(\pm 1) =
\frac{1}{2} (2k+1)(2k+2)$, meaning a non-uniform convergence as $x \to
0 $ or $x \to 1$ as well as the large-$n$ dominance of the end-point
behavior. In this limit we have \cite{RuizArriola:2004ui}
\begin{eqnarray}
\left( \frac{\alpha(Q) }{\alpha(Q_0)} \right)^{\gamma_n^{(0)}/(2
\beta_0)} \sim n^{-8 r(Q_0,Q)} e^{2(3-4\gamma)r(Q_0,Q)}.\nonumber \\
\end{eqnarray}  
Only even-$n$
terms contribute in Eq.~(\ref{eq:evolpda}), hence we impose this condition for
integer $n$ and then extend $n$ to real values. 
The summation in Eq.~(\ref{eq:evolpda}) is then replaced with an
integral,
\begin{eqnarray}
\phi(x,Q) &\sim& \frac12 8 x e^{(3/4-\gamma)a}\int_0^\infty \!\!\!\!\! dn \,C_n^\frac32 (2x-1) n^{-a-1}\, , \nonumber \\ 
\label{eq:pda-int}
\end{eqnarray} 
where the factor of $1/2$ comes from the summation over even $n$ only, and 
\begin{eqnarray}
a= 8 r (Q_0,Q). 
\end{eqnarray}
The Gegenbauer polynomials $C_n^\frac32 (\xi)$ in the variable \mbox{$\xi = 2 x-1$} satisfy
the differential equation~\cite{Abramowitz}
\begin{eqnarray}
(1-\xi^2) y''(\xi) - 4 \xi y'(\xi) + n(n+3) y(\xi) =0  \, , 
\end{eqnarray}
which upon the substitution 
\begin{eqnarray}
y(\xi) = \frac{u(\xi)}{1-\xi^2} \, , 
\end{eqnarray} 
transforms into a Schr\"odinger-like equation at zero energy,
\begin{eqnarray} 
u''(\xi) + \frac{n^2+3 n +2}{1-\xi^2} u(\xi) =0 \, .
\label{eq:sch} 
\end{eqnarray} 
Here the interval of interest, $-1 \le \xi \le 1 $, corresponds to the
classically allowed region and the potential is attractive. In the
large-$n$ limit the solution oscillates rapidly and a semiclassical WKB
approximation~\cite{GalindoPascual} might be used.  Here in order to
analyze the limit $x \to 0$ we consider the differential equation,
Eq.~(\ref{eq:sch}), which in the limit $x \to 0 $ and $n \gg 1 $
transforms into a zero-energy Coulomb-like problem.
Its solution can be generally written in terms of the Bessel functions, 
\begin{eqnarray}
C_n^\frac32 (2x-1) = \frac{c_1 J_1 ( 2 n \sqrt{x} ) + c_2 Y_1 ( 2 n\sqrt{x})}{\sqrt{x}}.
\end{eqnarray} 
where actually $c_2=0$, since the Gegenbauer polynomials are regular
at the end-point $x=0$. The undetermined constant $c_1$ may be
obtained by matching the small argument expansion $J_1(z) = z/2 +
\dots $ to the value $ C_{n}^{3/2}(1) \sim \frac{n^2}{2}$, yielding for
$n \gg 1$ and $x \to 0 $ the formula
\begin{eqnarray}
C_n^\frac32 (2x-1) \sim \frac{n}{2 \sqrt{x}} J_1 ( 2 n \sqrt{x})  \, .
\label{eq:bess}
\end{eqnarray} 
As we see in the low-$x$ and large-$n$ limit there is a scaling behavior
of the Gegenbauer polynomials, 
\begin{eqnarray}
C_n^\frac32 (x) \sim n F( n \sqrt{x}) /\sqrt{x} \, . 
\end{eqnarray} 
From Eq.~(\ref{eq:pda-int}) it is now clear that this
translates into the low-$x$ scaling behavior 
\begin{eqnarray}
\phi(x,Q) &\sim& x^{a/2}  e^{(3/4-\gamma)a} \int_0^\infty \, dt \, F(t) t^{-a} \, . 
\end{eqnarray} 
Evaluation of the integral and collecting the factors yields Eq.~(\ref{endPDA}).

%\bibliography{gpd}

\end{document}